\documentclass[lettersize,journal]{IEEEtran}
\usepackage{float}
\usepackage{amssymb}
\usepackage{bm}
\usepackage{amsthm, amsmath,amsfonts}
\usepackage[ruled, vlined]{algorithm2e}
\usepackage{array}
\usepackage{textcomp}
\usepackage{stfloats}
\usepackage{url}
\usepackage{verbatim}
\usepackage{graphicx}
\usepackage{subfigure}
\usepackage{cite}
\usepackage{booktabs}    
\usepackage{mathrsfs} 
\usepackage{setspace}
\usepackage{multirow}
\usepackage{amsmath} 
\allowdisplaybreaks[4] 
\usepackage{xcolor}
\usepackage{hyperref} 
\usepackage{color}
\usepackage{threeparttable}
\theoremstyle{definition}
\newtheorem{remark}{Remark}
\makeatletter
\renewcommand{\maketag@@@}[1]{\hbox{\m@th\normalsize\normalfont#1}}%
\makeatother

\makeatletter
\let\myorg@bibitem\bibitem
\def\bibitem#1#2\par{%
	\@ifundefined{bibitem@#1}{%
		\myorg@bibitem{#1}#2\par
	}{%
		\begingroup
		\color{\csname bibitem@#1\endcsname}%
		\myorg@bibitem{#1}#2\par
		\endgroup
	}%
}

\makeatother
\begin{document}


\title{
Chirp Delay-Doppler Domain Modulation Based Joint Communication and Radar for Autonomous Vehicles
}
\author{Zhuoran Li, Zhen Gao, Sheng Chen, \textit{Life Fellow}, \textit{IEEE}, Dusit Niyato, \textit{Fellow}, \textit{IEEE}, Zhaocheng Wang, \textit{Fellow}, \textit{IEEE},  George K. Karagiannidis, \textit{Fellow}, \textit{IEEE}
	\thanks{Z. Li and Z. Gao are with the School of Information and Electronics, Beijing Institute of Technology, Beijing 100081, China, and also with the School of Interdisciplinary Science, Beijing Institute of Technology, Beijing 100081, China	(e-mails: \mbox{lizhuoran2000@qq.com}, \mbox{gaozhen16@bit.edu.cn}). \textit{(Corresponding author: Zhen Gao.)}
	
	 Sheng Chen is with the School of Computer Science and Technology, Ocean University of China, Qingdao 266100, China (e-mail: \mbox{sqc@ecs.soton.ac.uk}).
	
	Dusit Niyato is with the College of Computing and Data Science, Nanyang Technological University, Singapore 639798, (e-mail: \mbox{dniyato@ntu.edu.sg}).
	
	 Z. Wang is with Beijing National Research Center for Information Science and Technology, Department of Electronic Engineering, Tsinghua University, Beijing 100084, China, and Z. Wang is also with Tsinghua Shenzhen International Graduate School, Shenzhen 518055, China (e-mail: \mbox{zcwang@tsinghua.edu.cn}).

	G. K. Karagiannidis is with the Department of Electrical and Computer Engineering, Aristotle University of Thessaloniki, 54124 Thessaloniki, Greece (e-mail: \mbox{geokarag@auth.gr}).	
	}
}


\maketitle
\begin{abstract}	
	This paper introduces a sensing-centric joint communication and millimeter-wave radar paradigm to facilitate collaboration among intelligent vehicles.
	We first propose a chirp waveform-based delay-Doppler quadrature amplitude modulation (DD-QAM) that modulates data across delay, Doppler, and amplitude dimensions.
	Building upon this modulation scheme, we derive its achievable rate to quantify the communication performance.
	We then introduce an extended Kalman filter-based scheme for four-dimensional (4D) parameter estimation in dynamic environments, enabling the active vehicles to accurately estimate orientation and tangential-velocity beyond traditional 4D radar systems.
	Furthermore, in terms of communication, we propose a dual-compensation-based demodulation and tracking scheme that allows the passive vehicles to effectively demodulate data without compromising their sensing functions.
	Simulation results underscore the feasibility and superior performance of our proposed methods, marking a significant advancement in the field of autonomous vehicles.
	Simulation codes are provided to reproduce the results in this paper: \href{https://github.com/LiZhuoRan0/2026-IEEE-TWC-ChirpDelayDopplerModulationISAC}{https://github.com/LiZhuoRan0}.
\end{abstract}
\vspace{-5mm}
\begin{IEEEkeywords}
	Millimeter-wave, integrated sensing and communications, joint communication and radar, autonomous vehicles.
\end{IEEEkeywords}
\vspace{-6mm}
\section{Introduction}\label{Sec_Introduction}
\vspace{-3mm}
\subsection{Prior Works}\label{S1.1}

  \IEEEPARstart{R}{adar} and communication technologies have paralleled each other in their seven decades of evolution, sharing many commonalities in hardware, waveform designs, and algorithms\cite{ref_SPM_FanLiu,refZXY,refWZW,refKMLIoTJ,refKMLTSP,refLXH}.
Their integration is seen as a leap forward for next-generation networks, offering synergy gains.
  Three primary waveform types underpin joint communication and radar (JCR), which are communication-centric design (CCD), sensing-centric design (SCD), and joint design (JD), with JD representing an ideal but challenging goal due to its novelty and complexity\cite{ref_JSTSP_AndrewZhang}.

  Orthogonal frequency division multiplexing (OFDM) is a widely considered waveform in CCD-based JCR.
OFDM can achieve high-speed data transmission while also supporting certain sensing functionalities\cite{ref_LZR}.
  However, in situations where only essential control data needs to be transmitted, rather than files, audio, or video, the high peak-to-average power ratio of OFDM becomes the main concern.
  In scenarios where there is a high demand for sensing, a lower requirement for communication rates, and where communication capability is an added benefit, the SCD is undoubtedly appealing.

 A representative SCD waveform is the chirp. Chirp exhibits simplicity in implementation, high tolerance to Doppler effects, and constant modulus characteristics, making it highly suitable for sensing applications.
  Chirp can be employed in both pulse radar and continuous wave radar\cite{ref_SPM_AutomotiveRadar}.
  Pulse radar systems, common in military and long-range surveillance, are characterized by their large size and high peak radiated power. They demand fast radio frequency switching to toggle between transmit and receive modes\cite{ref_TSP_MAJoRCom}. In contrast, continuous-wave radars are commonly employed in civil applications, with a typical example being the millimeter-wave (mmWave) radar used in automotive platforms \cite{ref_TI_AWR2243, ref_JSTSP_DingyouMa_IM}. These systems are designed under strict size, cost, and average-power constraints, which favors characteristics such as constant-modulus transmissions, lower peak power, and seamless transmit/receive integration.
The MAJoRCom system using pulse radar was proposed in \cite{ref_TSP_MAJoRCom}, where the carrier frequency of the pulses, the activated antennas for each pulse, and the correspondence between them are all variable. These variations enable data modulation.
  Similar to the MAJoRCom system, the FRaC system proposed in \cite{ref_JSTSP_DingyouMa_IM} is a continuous-wave radar platform that modulates information via a combination of carrier selection, antenna selection, and waveform permutation.
  However, it makes the assumption of a perfectly known channel. Moreover, the decoding algorithm employs the maximum likelihood method with high computational complexity, making its implementation challenging.

  Due to its low cost and powerful sensing capabilities, mmWave radar, particularly chirp or frequency modulated continuous wave (FMCW) based one, has become a key sensor in intelligent autonomous vehicle systems \cite{ref_SPM_AutomotiveRadar, ref_2025_CST, ref_22_dataset,ref_Network}.
  Among the three major sensors (camera, lidar and radar) in autonomous vehicle systems, the camera lacks speed sensing capabilities and, like lidar, it is also sensitive to the environment \cite{ref_SPM_Lidar}.
  This makes cost-effective mmWave radar systems indispensable \cite{ref_2025_arXiv, ref_JSTSP_mmWaveRadar}. 
  In the limited space of onboard mmWave radar systems, higher integration is required.
  Multiple-input multiple-output (MIMO) radar utilizes a small number of antennas to expand the array aperture, leading to an increase in angular resolution\cite{ref_SPM_MIMORadar}. 
   Notably, the 76--81~GHz frequency band has been allocated for JCR applications in autonomous vehicle systems, enabling both high-precision sensing and potential communication capabilities \cite{ref_ITU-R-M2057}.
  Nevertheless, the sensing capabilities of sensors on a single vehicle are limited, which has led to the development of collaborative localization methods among multiple vehicles \cite{ref_23WC_SLAM, ref_24TSP_MPNN}.
  Limited spectrum resources need efficient spectrum utilization. However, a major issue with existing mmWave radar systems is the inability to effectively reuse time-frequency (TF) resources\cite{ref_SPM_InterferenceMitigation}.
  If many vehicles employ mmWave radar systems, there will be severe mutual interference.
This creates an opportunity for an enhanced level of collaboration among vehicles, facilitating JCR in autonomous vehicle systems.
  The sensing process is practically continuous, often manifesting as a tracking process characterized by its parameters gradually changing and following a certain rule.
  Therefore, data can be modulated onto the parameters with additional offsets, and the receiver can demodulate the data by identifying these offsets.
  However, existing mmWave radar JCR systems, except some cognitive radar or adaptive waveform design, do not effectively utilize time series information for data modulation\cite{refCogRad,refAdapWav}.

  In the conventional communication domain, only long range (LoRa) technology utilizes the characteristic of chirp in Internet of Things \cite{ref_Access_LoRaFromTaoWithCode}.
  Compared to mmWave radar systems, LoRa technology operates at a lower frequency band with an extremely narrow bandwidth, within the industrial scientific medical bands, such as 433~MHz, 868~MHz, and 915~MHz.
Therefore, synchronization is sufficient and it does not require complex channel estimation, where data demodulation is achieved using a special dechirp process\cite{ref_Access_ImmuneDoppler}. 
Extending the idea of LoRa technology directly to mmWave encounters challenges associated with non-ideal hardware characteristics \cite{ref_TI_ChirpConfig_Tail} and issues arising from excessively high sampling rates.
Therefore, extending LoRa technology directly to mmWave is not trivial.
Some studies have attempted to integrate LoRa technology with sensing.
The authors of \cite{ref_TBC_Loradar} used the FMCW non-linearity effect in radio frequency circuits to obtain LoRa signals, but this method is not widely applicable.

  From the above review of the related work, we can draw the following issues and limitations of the existing literature. 
\begin{enumerate}
	\item Some existing SCD-based JCR schemes are not suitable for civilian use \cite{ref_TSP_MAJoRCom}, and some schemes targeting civilian applications make ideal assumptions and have high algorithm complexity \cite{ref_JSTSP_DingyouMa_IM}.
	
	\item The extension of LoRa technology from lower frequencies to the mmWave band faces challenges due to imperfect hardware and excessive sampling rates, making this extension difficult to achieve.
	
	\item Current JCR solutions hardly capture time series for data modulation on sensing parameters, such as delay and Doppler. Since time series information is available, it is possible to modulate the data on the sensing parameters. Clearly, leveraging time series information can be beneficial for JCR but the existing schemes have not utilized this vital information.
\end{enumerate}
\vspace{-6mm}
\subsection{Our Contributions}\label{S1.2}
  The above consideration motivates our current work.
  Specifically, to bestow communication functionalities upon mmWave radar systems, we propose a tracking-based data modulation and demodulation scheme, committed to preserving the fundamental sensing capabilities unimpaired and to facilitating seamless integration of communication and radar functionalities.
  Our main contributions are summarized as follows.
\begin{itemize}	
	
	\item By employing a ``listen-before-talk" regime, vehicles can exploit the mixing property of chirp signals to identify and select idle time-frequency resources for transmission. Within the selected TF resources, vehicles can modulate delay, Doppler, and amplitude data onto a single-frame chirp signal, known as delay-Doppler quadrature amplitude modulation (DD-QAM). Importantly, the same transmitted chirp signal can be simultaneously used for environment sensing by analyzing its reflections, thereby enabling joint communication and radar functionality. Moreover, the achievable rate of the proposed DD-QAM is evaluated.
	
	\item By exploiting the tracking process, we use relative movement to estimate the orientation. Based on the orientation and the radial-velocity, the tangential-velocity can also be estimated.
	Therefore, we propose an extended Kalman filter (EKF)-based scheme for five-dimensional (5D) parameter estimation.	
	The 5D parameters include delay (or distance), Doppler (encompassing both tangential and radial velocity), azimuth angle, and elevation angle, whereas conventional four-dimensional (4D) sensing typically excludes tangential velocity.
	Additionally, quasi-off-grid estimation of delay and Doppler can be obtained through the effective use of spectrum leakage, and true dynamic off-grid estimation can be obtained during the tracking process.
	
	\item By leveraging the time-series information from the tracking process, vehicles can extract data by computing the parameter difference between the prediction based on previous parameters and the estimation derived from the current observation.
	Subsequently, the current estimation results, excluding data, represent the true parameters of the targets and are utilized for subsequent tracking.
	Therefore, data can be modulated and demodulated without compromising the sensing functionality. We refer to this demodulation scheme as ``dual-compensation-based demodulation and tracking scheme''.
	The term ``dual-compensation'' refers to the need to compensate for target parameters during data demodulation and to remove modulated data during tracking.
\end{itemize}
\vspace{-4mm}
\subsection{Notations}\label{S1.3}
  Unless otherwise stated, normal-face letters, boldface lower-case letters, boldface uppercase letters and calligraphic letters denote scalar variables, column vectors, matrices and tensors, respectively.
 The transpose and conjugate operators are denoted by $(\cdot)^{\text T}$ and $(\cdot)^*$, respectively.
  $\textsf{j}=\sqrt{-1}$ is the imaginary unit, $\mathbb{R}$ and $\mathbb{C}$ are the sets of real-valued and complex-valued numbers, respectively.
  For vector $\mathbf{x}\! \in\! \mathbb{C}^{N\times 1}$, $\mathbf{x}[n]$ denotes the $n$-th element of $\mathbf{x}$, $\text{len}\left(\mathbf{x}\right)$ denotes the length of $\mathbf{x}$, and ${\bf{x}}[n](\theta)$ means extracting the element of ${\bf{x}}$ indexed by $[n]$, where $\theta$ is the argument of ${\bf{x}}[n](\theta)$.
  For matrix $\mathbf{X}\! \in\! \mathbb{C}^{N\times M}$, $\mathbf{X}[n,m]$ denotes the $(n,m)$-th element of $\mathbf{X}$, ${\bf{X}}[n,m](\theta)$ means extracting the element of ${\bf{X}}$ indexed by $[n,m]$ with argument $\theta$, and $\mathbf{X}[:,m_1:m_2]$ denotes the sub-matrix composed of the columns from the $m_1$-th column to $m_2$-th column of $\mathbf{X}$.
	For tensor $\mathcal{Y}\! \in\! \mathbb{C}^{N\times M\times K}$, $\mathcal{Y}[n,m,k]$ denotes its $(n,m,k)$-th element.
	$\mathcal{N}(\mu, \Sigma)$ denotes the Gaussian distribution with mean $\mu$ and covariance $\Sigma$.
  The magnitude of real- or complex-valued number $s$ is denoted by $|s|$. For real number $s$, $\left\lfloor s \right\rfloor$ represents the largest integer less than or equal to $s$, and $\left[\kern-0.15em\left[ {s} \right]\kern-0.15em\right]$ is the integer closest to $s$.
  $E[\cdot]$ denotes the expectation operation, $\partial(\cdot)$ is the first-order partial derivative operation and $\text{d}(\cdot)$ is the first-order total derivative operation.
  $\mathbf{1}_n$ denotes the vector of size $n\times 1$ with all the elements being 1.
  $\odot$ represents the Hadamard product, $\otimes$ represents Kronecker product, and $\circledast$ is the linear convolution operator, while $\mod$ denotes the modulus operation.
  $s_1 \& s_2$ denotes the logical AND operation between $s_1$ and $s_2$, and $x:y:z$ represents the sequence from $x$ to $z$ with spacing $y$.
  The operator $\text{Avg}({\mathbf{X}})$ takes the arithmetic average over multidimensional array ${\mathbf{X}}$, and the operator $\Leftarrow$ assigns the value of the variable on the right to the left. The speed of light is denoted as $c$, and $\Pi (t)$ is the unit rectangular window function, which equals to $1$ in the interval $[0,1)$ and $0$ otherwise, while $\delta(t)$ denotes the unit impulse function.
\vspace{-4mm}
\section{Overview of Proposed JCR Procedure}\label{sec_ProcedureAndFrame}
  Autonomous vehicles on the road need to sense the environment to ensure safety.
  Based on whether they actively transmit JCR signals, autonomous vehicles are classified into two types: active vehicle (AV) and passive vehicle (PV). AVs and PVs are equipped with the same JCR transceivers.
  The only difference between them lies in the roles each entity plays: the AV serves as the active sensing party, responsible for modulating data onto the signal, while the PV functions as the passive sensing party, tasked with demodulating data from the signal. These entities are capable of interchanging roles depending on actual needs.
  In addition, vehicles that are neither AVs nor PVs are referred to as ``other vehicles'', and they only serve as targets sensed by the AVs and PVs. These ``other vehicles'' can switch to AVs or PVs based on their own needs.

\begin{figure}[!t]
	\centering
	\includegraphics[width=2.5in]{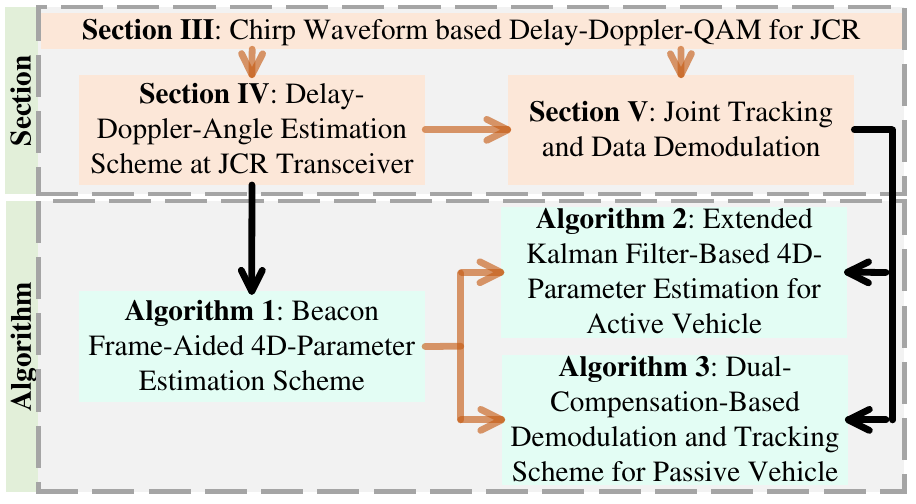}
	\vspace{-4mm}
	\caption{Organization of this paper.}
	\label{fig_FlowChart} 
	\vspace{-1mm}
\end{figure}

  The organization of this paper, namely, the proposed JCR procedure utilizing the chirp waveform-based on DD-QAM, is illustrated in Fig.~\ref{fig_FlowChart}.
  The AV transmits JCR signals and receives echoes to sense the environment, while the PV receives JCR signals to demodulate data and sense the environment.
  The proposed chirp waveform-based DD-QAM scheme, which modulates the data onto the delay, Doppler and complex amplitude of the transmit signal, is detailed in Section~\ref{sec_SignalModel}.
  Section~\ref{sec_ParaEst} covers parameter estimation techniques, and Section~\ref{sec_JointTrackDataDemod} contains target tracking and data demodulation schemes. 
  In Sections~\ref{sec_ParaEst} and \ref{sec_JointTrackDataDemod}, Algorithm~\ref{alg_ParaEst} summarizes the parameter estimation process, while Algorithm~\ref{alg_EKF_RT} details target tracking and data modulation of AV, and Algorithm~\ref{alg_EKF_CT} outlines target tracking and data demodulation of PV.
Algorithms~\ref{alg_EKF_RT} and \ref{alg_EKF_CT} both depend on the estimated parameters provided by Algorithm \ref{alg_ParaEst}.

\vspace{-6mm}
\section{Chirp Waveform based Delay-Doppler-QAM for JCR}\label{sec_SignalModel}
\vspace{-1mm}
For the proposed chirp waveform-based DD-QAM scheme, the data are modulated onto the delay, Doppler and complex amplitude of the transmit signal.
\begin{figure}[!t]
	\vspace{-3mm}
	\centering
	\includegraphics[width=3.5in]{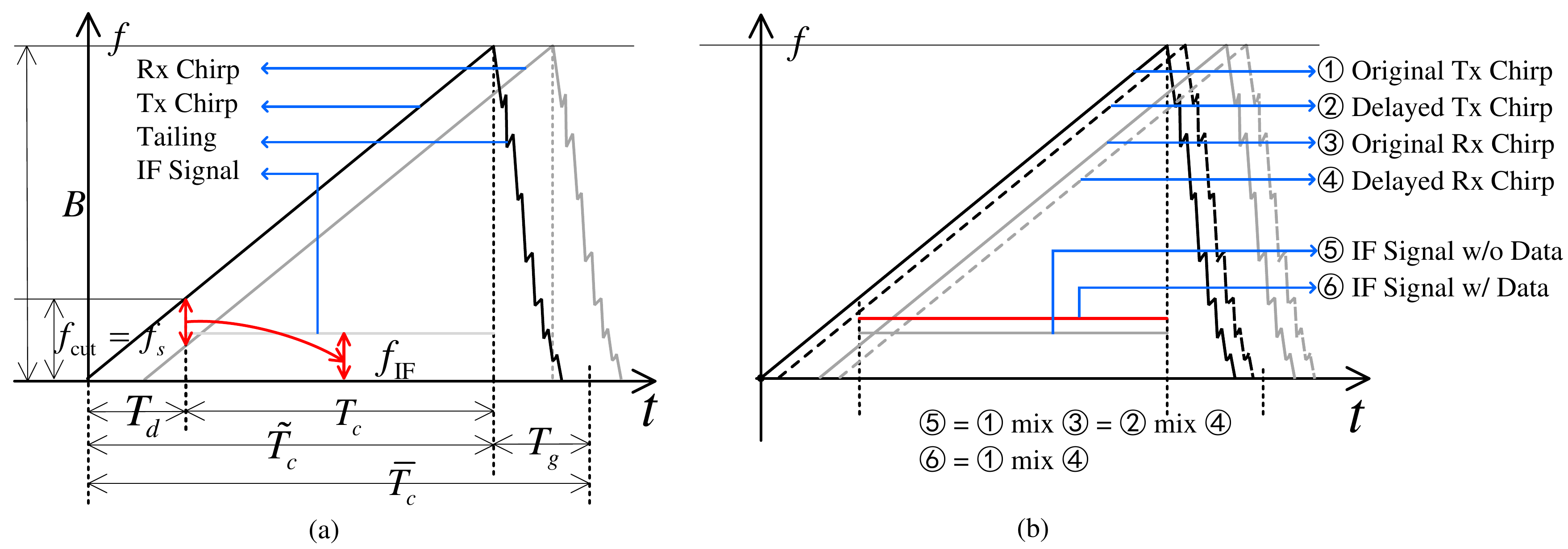}
	\vspace{-8mm}
	\caption{TF diagram used to illustrate the relationships among parameters and how data can be modulated in delay: (a)~basic dechirp process; (b)~data modulation and demodulation process, wherein the data is modulated in delay. The black chirp represents transmitted (Tx) one and the gray chirp represents the received (Rx) one.}
	\label{fig_chirp} 
	\vspace{-7mm}
\end{figure}
\vspace{-3mm}
\subsection{Signal Model under Single-path Single-Input Single-Output Channel}\label{sec_SISO}

  The transmitted single frame signal is defined as
\begin{align}\label{eqTxSig} 
	\vspace{-5mm}
	x(t) &= \big(\beta^{\text{(D)}}\big)^*\beta_{\text{tx}}  \sum\nolimits_{n_c=0}^{N_c-1} \Pi\left((t - n_c \bar{T}_c)/\tilde{T}_c\right) e^{-\textsf{j} 2\pi \frac{n_c}{N_c} f_v^{{\text{(D)}}}}\nonumber\\
	&\times e^{\textsf{j} s\big(t - n_c \bar{T}_c - \tau^{\text{(D)}}\big)},
	\vspace{-2mm}
\end{align}
where $\beta^{\text{(D)}}$ is the QAM symbol in one frame, $\beta^{\text{(D)}}$ together with $\beta_{\text{tx}}$ represents the amplitude of the transmitted signal, $N_c$ is the number of the chirps in one frame, $\tilde{T}_c$ is the chirp duration,
  $\tau^{\text{(D)}}$ is the data modulated on delay, $f_v^{\text{(D)}}$ is the data modulated on Doppler,
and $\bar{T}_c = \tilde{T}_c + T_g$ is the total duration of one unit of one frame containing the guard interval $T_g$, which is used to separate two adjacent chirps \cite{ref_TI_ChirpConfig_Tail}.
$T_c$ is the effective sampling time within one chirp duration $\tilde{T}_c$.
  The relationship among some parameters is shown in Fig.~\ref{fig_chirp}\,(a). The chirp's tailing effect is attributed to the non-ideal characteristics of the transmitter during the generation of high-frequency chirps.
  In Fig.~\ref{fig_chirp}, $f_{\text{cut}}$ denotes the cut-off frequency of the LPF, indicating that the intermediate-frequency (IF) signal will pass through the LPF before being sampled, with the cut-off frequency set equal to the sampling frequency $f_s$.
  $T_d=T_{\text{sen}}+T_{\text{com}}$ is the maximum delay containing the delay $T_{\text{sen}}$ for sensing and the delay $T_{\text{com}}$ for communication\footnote{Except for the physical-layer design, recent vehicular MAC-layer protocols reduce collision probability and coordinate time--frequency usage at the network/system level\cite{refC,ref_NR_V2X_Tutorial2021,refGPP}.}. For simplicity and without loss of generality, $T_{\text{sen}}\! =\! T_{\text{com}}\! =\! T_d/2$ is adopted. $T_d$ can be calculated as $f_s/B\! =\! T_d/\tilde{T}_c$. The received signal mixes with the locally generated \textit{origin signal} without modulated data and passes through an LPF with a cut-off frequency $f_{\text{cut}}\! =\! f_s$. The received signals with delay larger than $T_d$ are filtered out in the analog domain since the frequency after mixing exceeds $f_{\text{cut}}$.  Thus, as long as the time intervals of different chirps exceed a certain threshold, they can be separated by the LPF in the analog domain without mutual interference. Inspired by this property, different vehicles can be interleaved with $2T_d$ for interference mitigation.
  The chirp signal $s(t)$ is defined as
\begin{equation}\label{eqChiSig} 
	s(t) = 2\pi \left((f_c-B/2)t + \alpha t^2/2\right),
\end{equation}
where $f_c$ is the carrier frequency, $\alpha = B/\tilde{T}_c$ is the chirp rate, and $B$ is the bandwidth occupied by the chirp.
  The echo channel is defined as
  	\vspace{-1mm}
\begin{equation}\label{eqEcCh} 
	h(t) = \beta_h \delta\left( t - \tau (t) \right),
\end{equation}
where $\tau(t)=2(r+vt)/c$ is the delay of the echo channel, in which $r$ and $v$ are the distance and radial-velocity between the transmitter and the target, respectively, while $\beta_h$ is the channel gain.
  Therefore, the echo signal is defined as $y_{{\text{echo}}} = x(t) \circledast h(t) = \beta_h x\left(t-\tau(t)\right)$.

  Let us mix the \textit{original signal} $x_{\text {orig}}(t)= \sum\nolimits_{n_c=0}^{N_c-1} \Pi\left((t - n_c \bar{T}_c)/\tilde{T}_c\right) e^{\textsf{j} s\big(t - n_c \bar{T}_c \big)}$, which is not modulated by $\beta^{\text{(D)}}$, $\tau^{\text{(D)}}$ and $f_v^{\text{(D)}}$, with the echo signal $y_{{\text{echo}}}(t)$.
  Here ``mix'' refers to the multiplication of $x_{\text{orig}}(t)$ with the conjugate of $y_{{\text{echo}}}(t)$, i.e., $y(t)=x_{\text {orig}}(t) \cdot y^*_{{\text{echo}}}(t)$.
  We take only the part from $\tilde{T}_c-T_c + n_c\bar{T}_c$ to $\tilde{T}_c + n_c\bar{T}_c$ of the mixing results as\vspace{-1mm}
{\small\begin{align}\label{eq_RcvSig} 
	y(t) &= \tilde{\beta} \beta^{\text{(D)}} \sum\nolimits_{n_c=0}^{N_c-1}\Pi \left({\left(t-n_c\bar{T}_c-(\tilde{T}_c-{T_c})\right)/T_c}\right)\nonumber \\
	&\times e^{\textsf{j}2\pi \frac{n_c}{N_c} f_v^{{\text{(D)}}}} e^{\textsf{j} s \big(t-n_c\bar{T}_c\big) - \textsf{j} s \big(t-\tau(t) - n_c \bar{T}_c - \tau^{\text{(D)}}\big)} \nonumber \\
	&\mathop \approx \limits^1  \tilde{\beta} \beta^{\text{(D)}} \sum\nolimits_{n_c=0}^{N_c-1}\Pi \left({\left(t-n_c\bar{T}_c-(\tilde{T}_c-{T_c})\right)/T_c}\right) \nonumber \\
	&\times e^{\textsf{j}2\pi \frac{n_c}{N_c} f_v^{{\text{(D)}}}} e^{\textsf{j}2\pi\big(f_c(\tau(t)+\tau^{\text{(D)}})+\alpha (t-n_c\bar{T}_c)(\tau(t)+\tau^{\text{(D)}})\big)} \nonumber\\
	&= \tilde{\beta}\beta^{\text{(D)}} \sum\nolimits_{n_c=0}^{N_c-1} \Pi \left({\left(t-n_c\bar{T}_c-(\tilde{T}_c-{T_c})\right)/T_c}\right)\nonumber \\ 
	&\times e^{\textsf{j}2\pi \frac{n_c}{N_c} f_v^{{\text{(D)}}}} e^{\textsf{j}4\pi \big(f_c + \alpha (t-n_c\bar{T}_c)\big)\frac{r+r^{\text{(D)}}/2}{c}}  e^{\textsf{j}4\pi \big(f_c + \alpha (t-n_c\bar{T}_c)\big)\frac{v t}{c}},
\end{align}}
where $r^{\text{(D)}}=c \cdot \tau^{\text{(D)}}$, which can be regarded as another form of data modulated on delay, $T_c$ is the effective sampling time within one chirp duration $\tilde{T}_c$, and $\tilde\beta=\beta_{\text{tx}}\beta_{h}$ together with $\beta^{\text{(D)}}$ represents the amplitude of the received signal.
  The reason for the approximation $\mathop  \approx \limits^1$ is due to the fact that $\alpha(\tau(t)+\tau^{\text{(D)}})/2\ll f_c$ and $(\tau(t)+\tau^{\text{(D)}})/2\ll t$.
  The relationship among $\bar{T}_c,\ \tilde{T}_c,\ \text{and}\ T_c$ is shown in Fig. \ref{fig_chirp}\,(a). We omit the noise term in (\ref{eq_RcvSig}) to simplify the expression, and this simplification similarly applies to the received signal in the subsequent text. 
	
  Let $t=n_fT_s+n_c\bar T_c + \tilde{T}_c-{T_c}$, and the received signal after sampling is given by\vspace{-1mm}
{\small\begin{align}\label{eq_DiscreteRcvSig} 
	&y(t){|_{t = {n_f}{T_s} + {n_c}{{\bar T}_c} + \tilde{T}_c - T_c}} \approx \nonumber \\
	&\beta \beta^{\text{(D)}} e^{\textsf{j}2\pi n_f \frac{2\alpha T_s}{c}\big(r+\frac{r^{\text{(D)}}}{2}\big)} e^{\textsf{j}2\pi n_c \big(\frac{2 \bar{T}_c f_c}{c} v + \frac{f_v^{{\text{(D)}}}}{N_c}\big)} \overset{\Delta}{=} \mathbf{Y}[{n_f},{n_c}],
\end{align}}
where $\beta\! =\! \tilde{\beta} e^{\textsf{j}4\pi \big((f_c + \alpha (\tilde{T}_c - T_c)) \frac{r+\frac{r^{\text{(D)}}}{2}}{c} + \alpha (\tilde{T}_c - T_c) \frac{v(\tilde{T}_c - {T_c})}{c} \big)}\times$
$ e^{\textsf{j}4\pi f_c \frac{v(\tilde{T}_c - {T_c})}{c}}$, $\mathbf{Y}\! \in\! \mathbb{C}^{N_f\times N_c}$, $n_f\! \in\! \{0, 1, \cdots, T_c f_s-1\}$ is the index of the sampling points within one chirp, $f_s$ is the sampling frequency, $T_s=1/f_s$ is the sampling interval, and $N_f=T_cf_s$.
  We utilize $\alpha n_fT_s \ll f_c$, ${2v(\tilde T_c - {T_c}) + {{{f_c}v}}/{\alpha}} \ll r$ and $\alpha v(\tilde{T}_c - T_c)\ll f_c v$ to obtain (\ref{eq_DiscreteRcvSig}).
A conclusion similar to (\ref{eq_DiscreteRcvSig}) can be found in \cite{ref_JSTSP_DingyouMa_IM}.

\begin{figure}[!t]
	\centering
	\color{black}
	\includegraphics[width=3.5in]{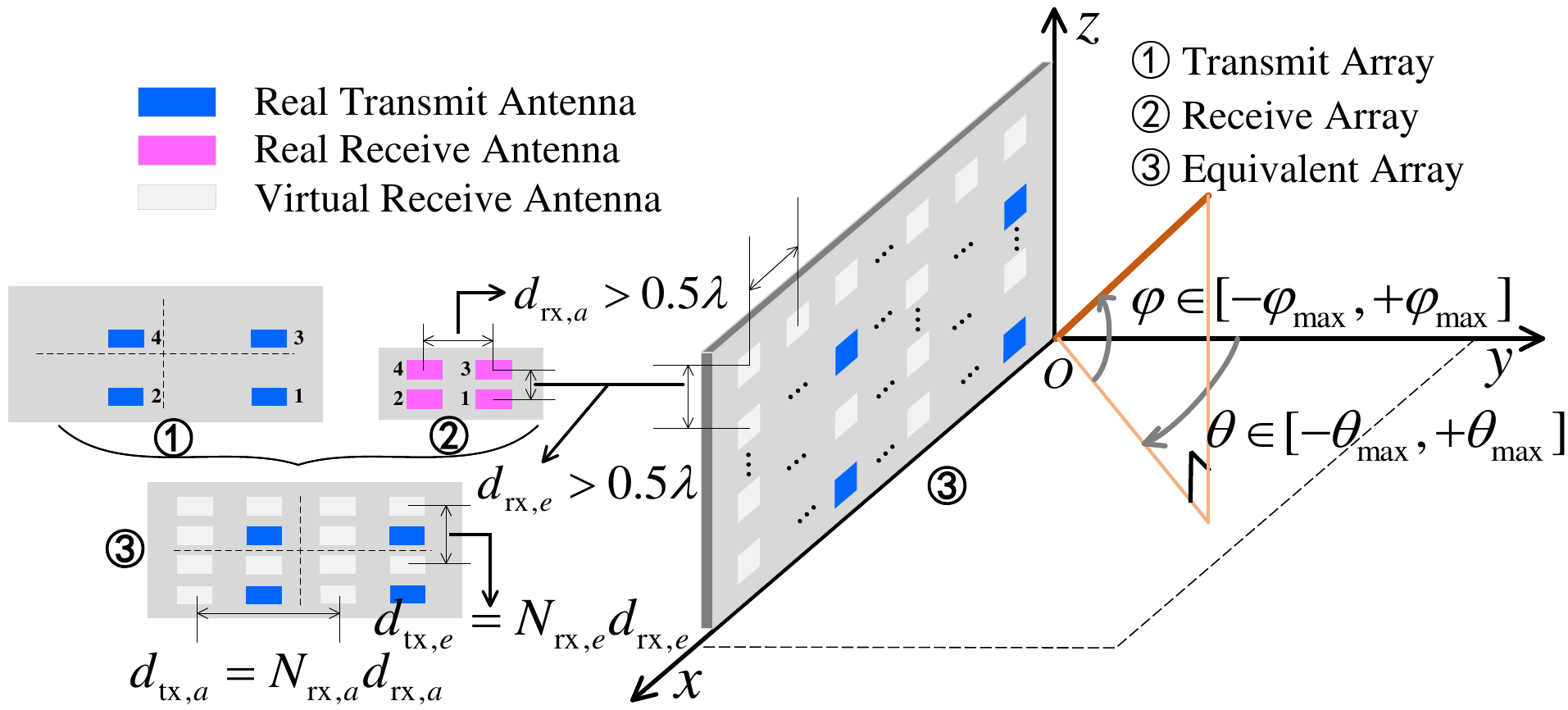}
	\vspace{-3mm}
	\caption{Schematic diagram of the transmit and receive array architecture.}
	\label{fig_array} 
	\vspace{-5mm}
\end{figure}

\vspace{-3mm}
\subsection{Signal Model under Multipath MIMO Channel}\label{S3.2}
\vspace{-1mm}
  As shown in Fig.~\ref{fig_array}, the transmit and receive arrays are placed on the $x$-$z$ plane. The position of each element in the transmit array (receive array) is defined by $\mathbf{p}_{\text{tx}}[n_{\text{tx}}]\! =\! [n_{\text{tx},a}d_{\text{tx},a},\ 0,\ n_{\text{tx},e}d_{\text{tx},e}]^{\text T}$ ($\mathbf{p}_{\text{rx}}[n_{\text{rx}}]\! =\! [n_{\text{rx},a}d_{\text{rx},a},\ 0,\ n_{\text{rx},e}d_{\text{rx},e}]^{\text T}$), where ``$n$'' is the index of the antenna starting from 0 and ``$d$'' is the antenna spacing, while subscript ``$\text{tx}$'' (``$\text{rx}$'') represents ``transmit'' (``receive''), and subscript ``$a$'' (``$e$'') stands for ``azimuth'' (``elevation''), which represents horizontal (vertical) direction.
  $N_{\text{tx}}\! =\! N_{\text{tx},a}N_{\text{tx},e}$ ($N_{\text{rx}}\! =\! N_{\text{rx},a}N_{\text{rx},e}$) is the number of transmit (receive) antennas arranged in $N_{\text{tx},a}$ ($N_{\text{rx},a}$) columns and $N_{\text{tx},e}$ ($N_{\text{rx},e}$) rows.
  Moreover, ${n_{\text{tx},a}}\! =\! {n_{\text{tx}}}\! \mod\! {N_{\text{tx},a}}$ (${n_{\text{rx},a}}\! =\! {n_{\text{rx}}}\! \mod\! {N_{\text{rx},a}}$), and ${n_{\text{tx},e}}\! =\! \left\lfloor {{n_{\text{tx}}}/{N_{\text{tx},a}}} \right\rfloor$ (${n_{\text{rx},e}}\! =\! \left\lfloor {{n_{\text{rx}}}/{N_{\text{rx},a}}} \right\rfloor$).

A frame of signal transmitted by the $n_{\text{tx}}$-th antenna is defined as\vspace{-1mm}
\begin{align}\label{eqMIMOs} 
	\mathbf{x}[n_{\text{tx}}](t) &= \big(\beta^{\text{(D)}}\big)^*\beta_{\text{tx}}  \sum\nolimits_{n_c=0}^{N_c-1} \Pi \left({(t-n_c\bar{T}_c)}/{\tilde{T}_c}\right) \mathbf{Q}[{n_{\text{tx}}},{n_c}] \nonumber \\
	&\times e^{-\textsf{j}2\pi \frac{n_c}{N_c} f_v^{\text{(D)}}} e^{\textsf{j}s\big(t-n_c\bar{T}_c-\tau^{\text{(D)}}\big)}.
\end{align}
Different values of ${\mathbf{Q}[{n_{\text{tx}}},{n_c}]}$ correspond to the different slow-time MIMO radar waveform orthogonality schemes\cite{ref_SPM_MIMORadar}.
  ${\mathbf{Q}[{n_{\text{tx}}},{n_c}]}$ allows the receiver to distinguish multiple transmit antennas and extend the vehicle's effective antenna aperture.
We adopt Doppler division multiplexing (DDM) scheme and  ${\mathbf{Q}[{n_{\text{tx}}},{n_c}]} = e^{\textsf{j}2\pi {n_c}{n_{\text{tx}}}/{N_{\text{tx}}}}$ \cite{ref_JSTSP_mmWaveRadar, ref_SPM_MIMORadar}.
The channel is defined as 
\begin{align}\label{eqMIMOc} 
	\mathbf{H}^{(\cdot )}[n_{{\text{tx}}},n_{{\text{rx}}}](t) &= \sum\nolimits_{l = 1}^{L^{( \cdot )}} \beta_{h,l}^{( \cdot )} e^{\textsf{j}2\pi \frac{f_c}{c} \big(\big({\bf{d}}_{{\text{tx}},l}^{({\text{A}})}\big)^{\text T} {\bf{p}}_{{\text{tx}}}[{n_{{\text{tx}}}}] + \big({\bf{d}}_{{\text{rx}},l}^{(\cdot )}\big)^{\text T} {\bf{p}}_{{\text{rx}}}[{n_{{\text{rx}}}}]\big)} \nonumber \\
	&\times \delta\big(t - \tau_l^{(\cdot )}(t)\big).
\end{align}
Here the superscript ``($\cdot$)'' is either ``(A)'' or ``(P)'' depending on whether it represents the AV's echo channel or the channel from the AV to the PV. 
  The index $l$ for the AV's echo channel $\mathbf{H}^{(\text{A})}[n_{{\text{tx}}},n_{{\text{rx}}}](t)$ indicates the $l$-th target and there are a total of $L^{({\text{A}})}$ targets, while the index $l$ for the channel from the AV to the PV $\mathbf{H}^{(\text{P})}[n_{{\text{tx}}},n_{{\text{rx}}}](t)$ indicates the $l$-th path and there are a total of $L^{({\text{P}})}$ paths\footnote{We assume that each target corresponds to a single, dominant line-of-sight path. This is a common and practical assumption in automotive radar research, as non-line-of-sight paths suffer from significant attenuation at millimeter-wave frequencies, rendering them unreliable for robust angle estimation\cite{ref_JSTSP_DingyouMa_IM}.}.
  The other terms in (\ref{eqMIMOc}) are now elaborated.
	
  The superscript of $\mathbf{d}_{\text{tx},l}$ is always ``(A)'' as the AV is the transmitter. For $\mathbf{H}^{(\text{A})}[n_{{\text{tx}}},n_{{\text{rx}}}](t)$ ($\mathbf{H}^{(\text{P})}[n_{{\text{tx}}},n_{{\text{rx}}}](t)$), $\mathbf{d}_{\text{tx},l}^{\text{(A)}}\! =\! \big[\cos \varphi_{\text{tx},l}^{\text{(A)}} \sin \theta_{\text{tx},l}^{\text{(A)}}, \cos \varphi_{\text{tx},l}^{\text{(A)}} \cos \theta_{\text{tx},l}^{\text{(A)}}, \sin \varphi_{\text{tx},l}^{\text{(A)}}\big]^{\text T}$ denotes the unit direction vector of the transmitter with respect to the $l$-th target (the $l$-th path), and $\mathbf{d}_{\text{rx},l}^{(\cdot)}\! =\! \big[\cos \varphi_{\text{rx},l}^{(\cdot)} \sin \theta_{\text{rx},l}^{(\cdot)} ,$ $\cos \varphi_{\text{rx},l}^{(\cdot)} \cos \theta_{\text{rx},l}^{(\cdot)} ,\ \sin \varphi_{\text{rx},l}^{(\cdot)}\big]^{\text T}$ is the unit direction vector of the receiver with respect to the $l$-th target (the $l$-th path), where $\theta_{\text{tx},l}^{(\cdot)}$, $\varphi_{\text{tx},l}^{(\cdot)}$, $\theta_{\text{rx},l}^{(\cdot)}$ and $\varphi_{\text{rx},l}^{(\cdot)}$ are the azimuth of departure (AoD), elevation of departure (EoD), azimuth of arrival (AoA) and elevation of arrival (EoA) with respect to the $l$-th target (the $l$-th path), respectively.
  For the echo channel, $\theta_{\text{tx},l}^{{\text{(A)}}}\! =\! \theta_{\text{rx},l}^{{\text{(A)}}}$ and $\varphi_{\text{tx},l}^{{\text{(A)}}}\! =\! \varphi_{\text{rx},l}^{{\text{(A)}}}$.
	
  The delay of the echo channel is $\tau_l^{({\text{A}})}(t)\! =\! 2\big(r_l^{({\text{A}})}\! +\! v_l^{({\text{A}})}t\big)/c$, where $r_l^{({\text{A}})}$ and $v_l^{({\text{A}})}$ are the distance and radial-velocity between the $l$-th target and the AV, respectively.
	The delay of the channel from the AV to PV is given by $\tau_l^{({\text{P}})}(t)=(r_l^{\text{(P)}}(t)+v_l^{\text{(P)}}t)/c$, where $r_l^{({\text{P}})}$ and $v_l^{({\text{P}})}$ represent the distance and the radial-velocity observed by the PV associated with the $l$-th path, respectively.
  Moreover, $v_l^{\text{(P)}}$ can be calculated according to the bi-static radar principle. Specifically, $r_l^{\text{(P)}}(t)=r_{\text{AS},l}(t)+r_{\text{PS},l}(t)$ and $v_l^{\text{(P)}}={\text{d}r_l^{\text{(P)}}(t)}/{\text{d}t}$, where $r_{\text{AS},l}(t)$ is the distance between the AV and the $l$-th scatterer, and $r_{\text{PS},l}(t)$ is the distance between the PV and the $l$-th scatterer. In the special case where the signal is transmitted from the AV to the PV directly, $v^{\text{(P)}}={\text{d} r_{\text{AP}}(t)}/{\text{d}t}$, where $r^{\text{(P)}}(t)=r_{\text{AP}}(t)$ is the distance between the AV and the PV.

	The schematic diagram of the transmit and receive array architecture is depicted in Fig.~\ref{fig_array}, where we set $d_{\text{tx},a}\! =\! N_{\text{rx},a}d_{\text{rx},a}$ and $d_{\text{tx},e}\! =\! N_{\text{rx},e}d_{\text{rx},e}$. Since the vehicle's field of view (FoV) is limited, $d_{\text{rx},a}$ and $d_{\text{rx},e}$ can be set to larger than $\lambda/2$ to increase the overall antenna aperture, where $\lambda$ is the wavelength.
  This setting leads to the increase of the angular resolution without increasing the number of antennas and introducing the angle ambiguity.
  If the FoVs are $\theta\! \in\! [-\theta_{\text{max}}, \theta_{\text{max}}]$ and $\varphi\! \in\! [-\varphi_{\text{max}}, \varphi_{\text{max}}]$, $d_{\text{rx},a}\! =\! \lambda/(2\sin \theta_{\text{max}})$ and $d_{\text{rx},e}\! =\! \lambda/(2\sin \varphi_{\text{max}})$.
	
  We mix the \textit{original signal} $\mathbf{x}_{\text{orig}}(t)\! =\! \sum\nolimits_{n_c=0}^{N_c-1} \Pi \left({(t\! -\! n_c\bar{T}_c)}/{\tilde{T}_c}\right) e^{\textsf{j}s\big(t-n_c\bar{T}_c\big)}$, which is not modulated by $\beta^{\text{(D)}}$, $\tau^{\text{(D)}}$ and $f_v^{\text{(D)}}$, with the echo signal.
Specifically, we mix ``$\textcircled{1}$'' with ``$\textcircled{4}$'' to obtain ``$\textcircled{6}$'' as illustrated in Fig.~\ref{fig_chirp}\,(b).
  The process to obtain the signal after mixing and sampling at the AV is similar to the process used to obtain the signal in (\ref{eq_DiscreteRcvSig}), and it can be expressed as
  \begin{align}\label{eq_RxSignalRT}	
	&\mathcal{Y}^{(\text{A})}[{n_f},{n_c},{n_{\text{rx}}}] \nonumber \\
	& =\sum\nolimits_{l = 1}^{L^{\text{(A)}}} \sum\nolimits_{{n_{\text{tx}}} = 0}^{{N_{\text{tx}}} - 1} \beta^{({\text{A}})}_l \mathbf{Q}[{n_{\text{tx}}},{n_c}] e^{\textsf{j}4\pi \frac{{{n_f}{T_s}}}{c}\alpha \big(r^{({\text{A}})}_l+\frac{r^{\text{(D)}}}{2}\big)} e^{\textsf{j}2\pi \frac{n_c}{N_c}f_v^{\text{(D)}}} \nonumber \\
	& \times \beta^{\text{(D)}} e^{\textsf{j}4\pi {f_c}\frac{{v^{({\text{A}})}_l{n_c}{{\bar T}_c}}}{c}} e^{\textsf{j}2\pi \frac{f_c}{c} \left(\big({\bf{d}}_{\text{tx},l}^{({\text{A}})}\big)^{\text T} {\bf{p}}_{\text{tx}}[{{n_{\text{tx}}}}] + {\big({\bf{d}}_{\text{rx},l}^{({\text{A}})}\big)}^{\text T}{\bf{p}}_{\text{rx}}[{{n_{\text{rx}}}}]\right)} \nonumber \\
	&= \sum\nolimits_{l = 1}^{L^{\text{(A)}}} \sum\nolimits_{{n_{\text{tx}}} = 0}^{{N_{\text{tx}}} - 1} \beta^{({\text{A}})}_l \beta^{\text{(D)}} e^{\textsf{j}2\pi \frac{{{n_f}}}{{{N_f}}}\left(f_{r,l}^{({\text{A}})}+f_r^{\text{(D)}}\right)} e^{\textsf{j}2\pi \frac{{{n_c}}}{{{N_c}}}\left(f_{v,l}^{({\text{A}})}+f_v^{\text{(D)}}\right)} \nonumber \\
	&\times \mathbf{Q}[{n_{\text{tx}}},{n_c}] e^{\textsf{j}2\pi \frac{{{f_c}}}{c}\left({{\left({\bf{d}}_{\text{tx},l}^{({\text{A}})}\right)}^{\text T}}{{\bf{p}}_{\text{tx}}[{{n_{\text{tx}}}}]} + {{\left({\bf{d}}_{\text{rx},l}^{({\text{A}})}\right)}^{\text T}}{\bf{p}}_{\text{rx}}[{{n_{\text{rx}}}}]\right)},
\end{align}
where $\mathcal{Y}^{\text{(A)}}\! \in\! \mathbb{C}^{N_f\times N_c\times N_{\text{rx}}}$, $\beta^{\text{(A)}}_l$ together with $\beta^{\text{(D)}}$ is the received complex amplitude of the $l$-th target's echo signal to the AV,
$f_{r,l}^{\text{(A)}}$ and $f_{v,l}^{\text{(A)}}$ are the normalized delay (or distance) frequency and normalized Doppler (or radial-velocity) frequency, respectively, which are given by
\begin{align}\label{eqAdi} 
	f_{r,l}^{\text{(A)}} &= 2 N_f T_s \alpha r^{\text{(A)}}_l/c \in [0, \, N_f), \nonumber \\
	f_{v,l}^{\text{(A)}} &= 2 N_c \bar{T}_c f_c v^{\text{(A)}}_l/c \in [0,\, N_c),
\end{align}
while ${f_r}^{({\text{D}})}\! =\! N_f T_s \alpha r^{\text{(D)}}/c\in\{0,1,\ldots,N_f-1\}$ is the data modulated in the delay. Note that $f_{r,l}^{\text{(A)}}+{f_r^{({\text{D}})}} < N_f$ should remain valid to ensure that $f_{r,l}^{\text{(A)}}<N_f$ and ${f_r^{({\text{D}})}}\leq N_f-1$.
  Additionally, ${f_v^{({\text{D}})}}\in\{0,1,\ldots,N_c-1\}$ or ${f_v^{({\text{D}})}}\in\{0,1,\ldots,N_c/N_{\text{tx}}-1\}$ is the data modulated in the Doppler depending on whether it is a beacon frame or a DDM frame.
The difference between a beacon frame and a DDM frame will be elaborated in Section~\ref{sec_DDMA}.
  If a beacon frame is used, only the first transmit antenna transmits. Owing to the cyclic shift characteristic of discrete Fourier transform (DFT), there is no constraint imposed on $f_{v,l}^{{\text{(A)}}} + {f_v}^{({\text{D}})}$, i.e., $f_{v,l}^{{\text{(A)}}} + {f_v}^{({\text{D}})}$ can be equal to or larger than $N_c$.

  We assume that the PV has the same signal $\mathbf{x}_{\text{orig}}(t)$ (``$\textcircled{1}$'' in Fig.~\ref{fig_chirp}\,(b)) as the AV, which can be used for mixing with the signal received at the PV (``$\textcircled{4}$'' in Fig.~\ref{fig_chirp}\,(b)).
  Therefore, the received signal after mixing and sampling at the PV can be obtained similarly to that of the AV in (\ref{eq_RxSignalRT}) as
  {\small\begin{align}\label{eq_CT_rcv}	
	&{\mathcal{Y}}^{({\text{P}})}[{n_f},{n_c},{n_{\text{rx}}}] \nonumber \\
	&= \sum\nolimits_{l = 1}^{L^{\text{(P)}}} \sum\nolimits_{{n_{\text{tx}}} = 0}^{{N_{\text{tx}}} - 1} \beta^{({\text{P}})}_l \mathbf{Q}[{n_{\text{tx}}},{n_c}] e^{\textsf{j}2\pi \frac{{{n_f}{T_s}}}{c}\alpha \left(r^{({\text{P}})}_l+r^{\text{(D)}}\right)} e^{\textsf{j}2\pi \frac{n_c}{N_c}f_v^{\text{(D)}}} \nonumber \\
	& \times \beta^{\text{(D)}} e^{\textsf{j}2\pi {f_c}\frac{{v^{({\text{P}})}_l{n_c}{{\bar T}_c}}}{c}} e^{\textsf{j}2\pi \frac{{{f_c}}}{c}\left({{\left({\bf{d}}_{\text{tx},l}^{({\text{A}})}\right)}^{\text T}}{\bf{p}}_{\text{tx}}[{{n_{\text{tx}}}}] + {{\left({\bf{d}}_{\text{rx},l}^{({\text{P}})}\right)}^{\text T}}{\bf{p}}_{\text{rx}}[{{n_{\text{rx}}}}]\right)} \nonumber \\
	&= \sum\nolimits_{l = 1}^{L^{\text{(P)}}} \sum\nolimits_{{n_{\text{tx}}} = 0}^{{N_{\text{tx}}} - 1} \beta^{({\text{P}})}_l \mathbf{Q}[{n_{\text{tx}}},{n_c}] e^{\textsf{j}2\pi \frac{{{n_f}}}{{{N_f}}}\left(f_{r,l}^{({\text{P}})}+f_r^{\text{(D)}}\right)} \beta^{\text{(D)}}\nonumber \\ 
	& \times  e^{\textsf{j}2\pi \frac{{{n_c}}}{{{N_c}}}\left(f_{v,l}^{({\text{P}})}+f_v^{\text{(D)}}\right)} e^{\textsf{j}2\pi \frac{{{f_c}}}{c}\left({{\left({\bf{d}}_{\text{tx},l}^{({\text{A}})}\right)}^{\text T}}{{\bf{p}}_{\text{tx}}[{{n_{\text{tx}}}}]} + {{\left({\bf{d}}_{\text{rx},l}^{({\text{P}})}\right)}^{\text T}}{\bf{p}}_{\text{rx}}[{{n_{\text{rx}}}}]\right)} ,	
\end{align}}
where $\mathcal{Y}^{\text{(P)}}\! \in\! \mathbb{C}^{N_f\times N_c\times N_{\text{rx}}}$, and\vspace{-1mm}
{\small\begin{align}\label{eqY(D)} 
	f_{r,l}^{\text{(P)}} &= N_f T_s \alpha r^{\text{(P)}}_l/c \in[0,\, N_f/2),\nonumber\\
	f_{v,l}^{\text{(P)}} &= N_c\bar{T}_c f_c v^{\text{(P)}}_l/c \in [0, \,N_c/2) ,	
\end{align}}
while $f_{r,l}^{\text{(P)}}$ and $f_{v,l}^{\text{(P)}}$ represent the normalized delay (or distance) frequency and normalized Doppler (or radial-velocity) frequency, respectively. Similarly, $f_{r,l}^{\text{(P)}}+{f_r^{({\text{D}})}}<N_f$ should be satisfied.

\begin{remark}\label{label_delay} 
	For ease of description and without loss of generality, we let $f_{r,l}^{\text{(A)}}\in[0,\, N_f/2)$, $f_{r,l}^{\text{(P)}}\in[0,\, N_f/2)$ and ${f_r^{({\text{D}})}}\in\{0,1,\ldots ,N_f/2-1\}$.
	Therefore, $f_{r,l}^{\text{(A)}}+{f_r^{({\text{D}})}}<N_f$ and $f_{r,l}^{\text{(P)}}+{f_r^{({\text{D}})}}<N_f$ always hold true.
	\hfill\qedsymbol
\end{remark}

\begin{remark}\label{R2}
	We term this modulation scheme DD-QAM, since data is modulated on delay, Doppler and complex amplitude of the transmitted signal.
	The rationale of DD-QAM lies in that temporal information in the tracking process can provide a prior dynamic and finite range for the parameters to be estimated.
	As long as the data modulated onto the parameters exceeds a certain distance, these parameters can be estimated, while simultaneously achieving data transmission.
	Note that the works in \cite{ref_TSP_MAJoRCom} and \cite{ref_JSTSP_DingyouMa_IM} adopted the inter-pulse modulation while our scheme adopts the inter-frame modulation.
	This distinction arises from our adoption of the DDM MIMO radar waveform orthogonality scheme to distinguish different transmit antennas at the receiver.
	Employing DDM for data modulation per chirp proves infeasible, resulting in a lower bit rate.
	By contrast, time-division multiplexing (TDM) does not have the aforementioned issues.
	TDM has the potential to enhance the bit rate by a factor of $N_c$, where $N_c$ is the number of chirps in one frame.
	However, a drawback is that only one antenna is active at any given time, which results in an SNR loss of $10\log_{10}(N_{\text{tx}})$ dB.
	As TDM can be viewed as a simplified and direct extension of DDM, we will not elaborate on it here.
	Nevertheless, its performance will be evaluated in the simulation section for comparison.
	
	The main distinction between our DD-QAM scheme and the existing chirp-based schemes lies in the domain used for data modulation.
	Our DD-QAM leverages the unique transmission and reception processes of mmWave radar systems \cite{ref_TI_AWR2243}.
	In this scheme, data is directly modulated onto the sensing parameters, facilitating seamless integration of communication and sensing, while also enabling low-complexity sensing and demodulation algorithms.
	In contrast, the scheme proposed in \cite{ref_TSP_MAJoRCom,ref_JSTSP_DingyouMa_IM} modulates data via antenna and frequency selection, necessitating the maximum likelihood algorithm at the receiver for demodulation, which significantly increases implementation complexity. 
	\hfill\qedsymbol
\end{remark}

\begin{figure}[!t]
\centering
\includegraphics[width=3in]{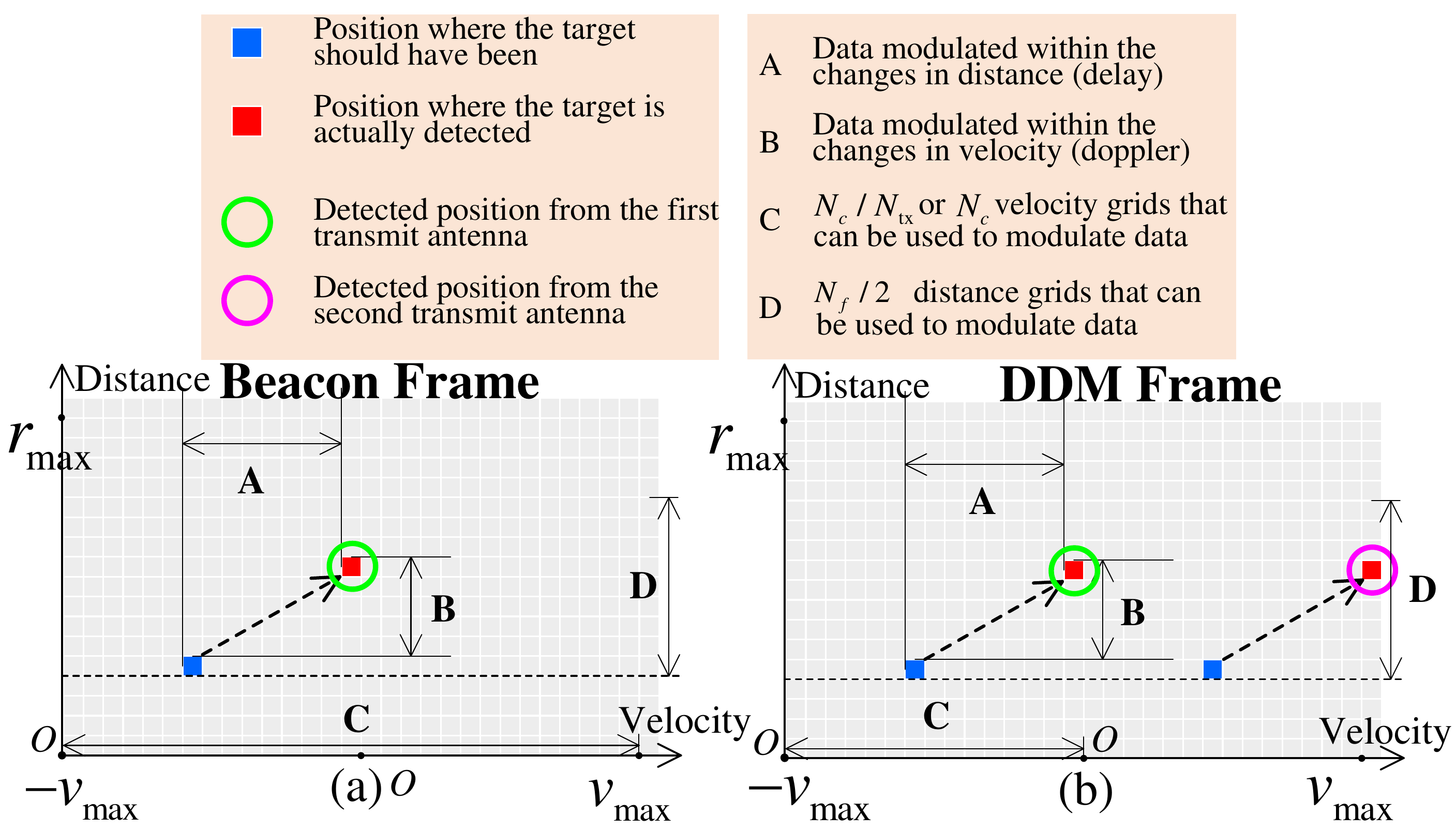}
\vspace{-4mm}
\caption{Schematic diagram illustrating the data modulation scheme of DD-QAM:
	(a) beacon frame; (b) DDM frame.
	This RDM is presented from the perspective of the AV or PV.
	It depicts two transmit antennas and one target in a receive antenna via DDM waveform orthogonality.
	It shows the way to modulate data in the variation of distance and velocity.
	The data modulated in complex amplitude is hidden in the complex amplitude of the detected position.}
\label{fig_RDM} 
\vspace{-4mm}
\end{figure}
\vspace{-5mm}
\subsection{Proposed Beacon Frame Aided DDM for DD-QAM}\label{sec_DDMA}

  DDM is a MIMO radar waveform orthogonality scheme. However, it reduces the maximum detection velocity by a factor of $N_{\text{tx}}$ \cite{ref_SPM_MIMORadar}.
  To address this issue, we propose the beacon frame aided DDM, which preserves the maximum detection velocity.
It is important to note that beacon frame aided DDM and DD-QAM are two techniques operating in different dimensions. 
  The former is independent of communication and helps the receiver to distinguish multiple transmit antennas, and it extends the vehicle's effective antenna aperture using ${\mathbf{Q}[{n_{\text{tx}}},{n_c}]}$.
  In contrast, the latter is a modulation scheme.

We can obtain the received signal after DFT at each receive antenna as \vspace{-1mm}
{\small\begin{align}\label{eq_2D_DFT}
	 \mathcal{\underline Y}^{(\cdot)}[{m_f},{m_c},n_{\text{rx}}]& = \sum\nolimits_{{n_c} = 0}^{{N_c} - 1} \sum\nolimits_{{n_f} = 0}^{{N_f} - 1} \mathcal{Y}^{({\cdot})}[{n_f},{n_c},{n_{\text{rx}}}] \nonumber\\
	&\times \mathbf{w}_f[{n_f}] \mathbf{w}_c[{n_c}] e^{-\textsf{j}2\pi \frac{{{m_f}{n_f}}}{{{N_f}}}} e^{-\textsf{j}2\pi \frac{{{m_c}{n_c}}}{{{N_c}}}},
\end{align}}
where $\mathcal{\underline{Y}}^{(\cdot)}\! \in\! \mathbb{C}^{N_f \times N_c \times N_{\text{rx}}}$ and superscript ``($\cdot$)'' is either ``(A)'' or ``(P)'' depending on whether it is the received signal of the AV or the PV, while
 ${\mathbf{w}_f}\! \in\! \mathbb{C}^{N_f}$ and ${\mathbf{w}_c}\! \in\! \mathbb{C}^{N_c}$ are the fast-time and slow-time window functions to mitigate the impact of the spectrum leakage.
  Then, the range-Doppler map (RDM) before the constant-false-alarm-rate (CFAR) detection, ${\bf{Y}}_{\text{RDM}}^{({\cdot})}\! \in\! \mathbb{C}^{N_f\times N_c}$, can be obtained as \vspace{-1mm}
\begin{equation}\label{eq_RDM} 
	{\bf{Y}}_{{\text{RDM}}}^{({\cdot})}[n_f,n_c] = \sum\nolimits_{{n_{\text{rx}}} = 0}^{{N_{\text{rx}}} - 1} {\left|{{\mathcal{\underline Y}}^{({\cdot})}}[{n_f},{n_c},{n_{\text{rx}}}]\right|/N_{\text{rx}}} ,
\end{equation}
to improve robustness. The RDM before detection is depicted in Fig.~\ref{fig_RDM}.
  Employing the CFAR detection on ${\bf{Y}}_{\text{RDM}}^{({\cdot})}$ yields the detection result ${\bf{Y}}_{\text{RDM,D}}^{({\cdot})}\! \in\! \mathbb{C}^{N_f\times N_c}$, whose element ${\bf{Y}}_{\text{RDM,D}}^{({\cdot})}[n_f,n_c]$ takes the value of 1 or 0, to indicate whether there is a target or not in the grid $[n_f,n_c]$.
  To address both velocity de-ambiguity and angle estimation \cite{ref_SPM_MIMORadar}, we use two frames to obtain all 4D parameters.
  The first frame serves as a beacon frame to mitigate the reduction in the maximum detected velocity. The second DDM frame serves the purpose of distinguishing different transmit antennas at the receiver.
  Specifically, only the first transmit antenna is activated in the beacon frame, ensuring that each target corresponds to a single detected region in the RDM.
  Therefore, the distance and velocity of each target can be obtained uniquely, as shown in Fig. \ref{fig_RDM}\,(a).
  All transmit antennas are then activated in the subsequent DDM frame, resulting in each target corresponding to $N_{\text{tx}}$ detected regions in the RDM.
  Despite potential changes in distance and velocity between the adjacent beacon frame and DDM frame, they can still be recognized due to their minimal variations, which will be verified through simulations.
  The values of the corresponding detected regions on ${\bf{Y}}_{\text{RDM}}^{({\cdot})}$ can be used to determine the azimuth and elevation angles, as shown in Fig.~\ref{fig_RDM}\,(b).

 If there is no modulated data, the position detected on the RDM in Fig.~\ref{fig_RDM} should be within the blue square area. In the presence of data modulation, the detected position is within the red square area.
  Since the PV has relatively accurate channel state information during tracking, it can demodulate the data modulated on delay and Doppler by distinguishing the difference between the red and blue squares. The maximum index that can be modulated onto the delay is $N_f/2-1$, as described in Remark~\ref{label_delay}.
  For the beacon frame, the maximum detectable Doppler corresponds to the maximum data that can be modulated on the Doppler. The difference in the maximum data modulated on Doppler and delay arises from the cyclic shift characteristic of the DFT in the Doppler domain, whereas any signal whose delay beyond the maximum detectable delay will be filtered out by the LPF.
  It should be noted that since $N_{\text{tx}}$ antennas transmit signals simultaneously in the DDM frame, the maximum data that can be modulated on the Doppler in the DDM frame is the maximum detectable Doppler divided by $N_{\text{tx}}$.

\begin{table}[!t]
\vspace{-2mm}
\centering
\begin{scriptsize}
\begin{threeparttable}
	\begin{scriptsize}
	\caption{Resolution and Range\textit{} of the 4D parameters}
	\label{table_SensingBound}
	\begin{tabular}{l|c|c}
		\hline \hline
		\multirow{1}{*}{\textbf{Parameters [unit]}} & \multicolumn{1}{c|}{\textbf{Resolution}}                    & \multicolumn{1}{c}{\textbf{Range}}  
		\\ \hline \hline
		Distance [m]                       &\Large$\frac{{c{{\tilde T}_c}}}{{2B{T_c}}}$&{\begin{tabular}[c]{@{}c@{}}$(0, \frac{{c{T_c}{f_{\text{cut}}}}}{{2(B - {f_{\text{cut}}})}})$\\ $=(0, \frac{{c(\tilde{T}_c-T_c)}}{2})$\end{tabular}} \\ \hline
		{\begin{tabular}[l]{@{}l@{}}Radial-\\ Velocity [m/s]\end{tabular}}                    &\Large $\frac{c}{{2N_c{{\bar T}_c}{f_c}}}$&{\begin{tabular}[c]{@{}c@{}}$( - 1,1) \times \min ($\\ $\frac{c}{{4{{\bar T}_c}{f_c}}},\frac{{c{{\tilde T}_c}}}{{4B{T_c}{N_c}{{\bar T}_c}}})$\end{tabular}}     \\ \hline
		sin(EoA) [NA]                   &$\frac{\lambda }{{{N_{\text{tx},e}}{N_{\text{rx},e}}{d_{\text{rx},e}}}}$&$(-\sin \varphi_{\text{max}}, \sin \varphi_{\text{max}})$  \\ \hline
		sin(AoA) [NA]                     &$\frac{\lambda }{{{N_{\text{tx},a}}{N_{\text{rx},a}}{d_{\text{rx},a}}\cos \varphi }}$&$(-\sin \theta_{\text{max}}, \sin \theta_{\text{max}})$   \\ \hline \hline
	\end{tabular}
\end{scriptsize}

\begin{scriptsize}
  \begin{tablenotes}
	 \item[1)] We use $r^{(\cdot)}_{\text{res}}$ and $v^{(\cdot)}_{\text{res}}$ to represent distance and velocity resolution, respectively, where superscript ``($\cdot$)'' is either ``(A)'' or ``(P)'' depending on whether it is the received signal from the AV or the PV.
	\item[2)] This table is for the AV. The resolution and range of distance and radial-velocity for the PV are twice those of the AV. The EoA resolution and AoA resolution of the PV are $\lambda/({{{N_{\text{rx},e}}{d_{\text{rx},e}}}})$ and ${\lambda }/({{{N_{\text{rx},a}}{d_{\text{rx},a}}\cos \varphi }})$, respectively. The EoA range and AoA range of the PV are the same with those of the AV.\begin{flushleft}
				\end{flushleft}
	\end{tablenotes}
\end{scriptsize}
\end{threeparttable}
\vspace{-8mm}
\end{scriptsize}
\end{table}
\vspace{-4mm}
\subsection{Resolution, Range and Rate of Chirp Waveform based DD-QAM}\label{S3.4}

  The resolution and range of the distance, radial-velocity and angle can be obtained through trigonometric relationships illustrated in Fig.~\ref{fig_chirp} and basic Fourier transform property \cite{ref_TI_ChirpConfig_Tail}.
  Due to space limitations, the detailed derivation is omitted here but the results are summarized in Table~\ref{table_SensingBound}.
  In general, $f_s \ll B$. Since the delay in the AV's received signal is caused by two-way distance while the delay in the PV's received signal is caused by only one-way distance, the resolution and range of the distance and velocity at the PV side are twice of those at the AV side.

  The number of bits that can be transmitted in one frame is \vspace{-1mm}
\begin{equation}\label{eq_order} 
	N_b = \left\lfloor {{{\log }_2}({N_f}/2)} \right\rfloor + \left\lfloor {{{\log }_2}({N_Q})} \right\rfloor  + {N_v},
\end{equation}
where ${N_v} = \left\lfloor {{{\log }_2}({N_c})} \right\rfloor$ or ${N_v} = \left\lfloor {{{\log }_2}({N_c/N_{\text{tx}}})} \right\rfloor$ depending on whether it is a beacon frame or a DDM frame, 
and $N_Q$ is the order of QAM. Thus, the bit rate is $N_b/(N_c\bar{T}_c)$ bits per second.
As for TDM, the number of bits that can be transmitted in one frame is $N_{b,\text{TDM}} = \left\lfloor {{{\log }_2}({N_f}/2)} \right\rfloor N_c + \left\lfloor {{{\log }_2}({N_Q})} \right\rfloor N_c$, where the bit rate is $\left(\left\lfloor {{{\log }_2}({N_f}/2)} \right\rfloor + \left\lfloor {{{\log }_2}({N_Q})} \right\rfloor \right)/\bar{T}_c$ bits per second.

\vspace{-4mm}
\section{Achievable Rate of DD-QAM}\label{Sec_Rate}
In this section, we derive the achievable rate of the proposed DD-QAM.
To facilitate the derivation, we reformulate the model from Section~\ref{sec_SignalModel} into a more compact matrix input-output form.
As MIMO is a direct extension of the SISO case, we illustrate the derivation using the SISO setting for clarity.

We define $\mathbf{X}_{\text{DD}}\in\mathbb{C}^{N_f\times N_c}$ as transmitted delay-Doppler-domain signal, and the received frequency-time-domain signal $\mathbf{Y}_{\text{FT}}\in\mathbb{C}^{N_f\times N_c}$ is given by
\begin{align}
	\mathbf{Y}_{\text{FT}} = \sum\nolimits_{l=1}^{L^{\text{P}}}
	\beta_{h,l}^{(\text{P})} \boldsymbol{\Lambda}_{\text{F},l} \mathbf{F}_{N_f}^{\text{H}} \mathbf{X}_{\text{DD}}  \mathbf{F}_{N_c}^{\text{H}} \boldsymbol{\Lambda}_{\text{T},l}
	+ \mathbf{W}_{\text{FT}},
\end{align}
where $\boldsymbol{\Lambda}_{\text{F},l}=\text{diag}(\boldsymbol{\lambda}_{\text{F},l})$ and $\boldsymbol{\Lambda}_{\text{T},l}=\text{diag}(\boldsymbol{\lambda}_{\text{T},l})$ are the $l$-th single path frequency-domain channel and time-domain channel, respectively.
$\text{diag}(\cdot)$ is the diagonalization operator.
$\boldsymbol{\lambda}_{\text{F},l} \in \mathbb{C}^{N_f}$ and $\boldsymbol{\lambda}_{\text{T},l} \in \mathbb{C}^{N_c}$ are single-tone signals.
$\mathbf{F}_{N_f} \in \mathbb{C}^{N_{f}\times N_f}$ and $\mathbf{F}_{N_c} \in \mathbb{C}^{N_{c}\times N_c}$ are Fourier transform matrices.
$\mathbf{W}_{\text{FT}} \in
\mathbb{C}^{N_f\times N_c}$ is white Gaussian noise.
We then can transform the frequency-time-domain received signal $\mathbf{Y}_{\text{TF}}$ to the delay-Doppler-domain received signal $\mathbf{Y}_{\text{DD}}\in\mathbb{C}^{N_f\times N_c}$ as
\begin{equation}
	\mathbf{Y}_{\text{DD}}
	= \mathbf{F}_{N_f} \mathbf{Y}_{\text{FT}} \mathbf{F}_{N_c}.
\end{equation}
By vectorizing $\mathbf{Y}_{\text{DD}}$, we can obtain
\begin{equation}
	\mathbf{y}_{\text{DD}}
	= \mathbf{H}_{\text{DD}}  \mathbf{x}_{\text{DD}} + \mathbf{w}_{\text{DD}},
\end{equation}
where $\mathbf{H}_{\text{DD}} = \sum_{l=1}^{L^{\text{P}}} \beta_{h,l}^{(\text{P})}
\left( \mathbf{F}_{N_c}^{\text{H}} \boldsymbol{\Lambda}_{\text{T},l} \mathbf{F}_{N_c} \right)^\text{T}
\otimes
\left( \mathbf{F}_{N_f} \boldsymbol{\Lambda}_{\text{F},l} \mathbf{F}_{N_f}^{\text{H}} \right)\in\mathbb{C}^{N_fN_c\times N_fN_c}$, $\mathbf{x}_{\text{DD}} = \text{vec}(\mathbf{X}_{\text{DD}})\in\mathbb{C}^{N_fN_c}$, $\mathbf{y}_{\text{DD}} = \text{vec}(\mathbf{Y}_{\text{DD}})\in\mathbb{C}^{N_fN_c}$, and $\mathbf{w}_{\text{DD}} = \text{vec}\left( \mathbf{F}_{N_f} \mathbf{W}_{\text{FT}} \mathbf{F}_{N_c} \right)\in\mathbb{C}^{N_fN_c}$.
$\mathbf{x}_{\text{DD}}$ contains only one non-zero element, which is a QAM symbol, and its position follows a uniform distribution.
It is worth noting that an additional constraints on $\mathbf{x}_{\text{DD}}$ is that the non-zero element in the delay-domain can only be located at indices $\{0,1,\ldots,N_f/2-1\}$.
Moreover, the maximum normalized frequencies of the corresponding single-tone signals $\boldsymbol{\lambda}_{\text{F},l}$ and $\boldsymbol{\lambda}_{\text{T},l}$ are $\tfrac{N_f/2-1}{N_f}$ and $\tfrac{N_{\text{tx}}-1}{N_c}$, respectively.
This is due to the reason stated in Remark \ref{label_delay} and the adopted DDM scheme.

To facilitate the subsequent derivations, we define the total number of possible codewords for
$\mathbf{x}_{\text{DD}}$ as $M=N_f/2\times N_c/N_{\text{tx}} \times N_Q$.
The $m$-th basis symbol of $\mathbf{x}_{\text{DD}}$ is denoted as $\mathbf{x}_{\text{DD},m}$.
Accordingly, for a fixed channel, the corresponding codeword is given by $\mathbf{c}_{m}=\mathbf{H}_{\text{DD}}\mathbf{x}_{\text{DD},m}$ and the codeword difference is defined as $\Delta_{m,m'}=\mathbf c_m-\mathbf c_{m'}$.
The additive noise is modeled as $\mathbf w_{\text{DD}}\sim\mathcal{CN}(\mathbf 0,\sigma^{2}\mathbf I_N)$.
The index of the codeword is denoted by $s\in\{1,\dots,M\}$.
Thus, the transmission model can be written as
\begin{equation}
	\mathbf y=\mathbf c_s+\mathbf w_{\text{DD}}.
\end{equation}
Defining conditional entropy as $h({\mathbf{y}}|s)$ and the entropy of the received signal as $h({\mathbf{y}})$, the achievable rate can be computed through mutual information as
\begin{align}\label{eq_mi}
	I(s;{\mathbf{y}}) &= h({\mathbf{y}}) - h({\mathbf{y}}|s) \nonumber \\
	&= {E_{s,{\mathbf{y}}}}\left(\ln \left( p({\mathbf{y}}|s)/p({\mathbf{y}})\right)\right)/\ln 2, 
\end{align}
where
\begin{equation}
	p(\mathbf y\mid s=m) =1/{(\pi\sigma^{2})^{M}}\; e^{-\|\mathbf y-\mathbf c_m\|^{2}/\sigma^{2}}
\end{equation}
and 
\begin{align}
	p(\mathbf y) & =1/M\sum\nolimits_{m'=1}^{M} p(\mathbf y\mid S=m')\nonumber \\
	& =(\pi\sigma^{2})^{-M}/M \sum\nolimits_{m'=1}^{M}\! \exp\bigl(-\|\mathbf y-\mathbf c_{m'}\|^{2}/\sigma^{2}\bigr).
\end{align}
Therefore, $h({\mathbf{y}}|s = m)$ can be obtained as 
{\small\begin{align}
	h({\mathbf{y}}|s = m)& =  - \int {p({\mathbf{y}}|s = m){\mkern 1mu} \ln p({\mathbf{y}}|s = m){\mkern 1mu} {\text{d}}{\mathbf{y}}}  \nonumber \\
	& = M\ln (\pi {\sigma ^2}) + M . 
\end{align}}
Since $h(\mathbf y | s)= \tfrac{1}{M}\sum_m h(\mathbf y | s=m)$, we can obtain $h(\mathbf y|s) = M\ln(\pi\sigma^{2}) + M$.
Furthermore, we can derive $h({\mathbf{y}})$ as follows:
{\small\begin{align}
	h({\mathbf{y}}) 
	&=  - \int {p({\mathbf{y}}){\mkern 1mu} \ln p({\mathbf{y}}){\mkern 1mu} {\text{d}}{\mathbf{y}}}  \nonumber \\
	&=  - \tfrac{1}{M}\sum\nolimits_k {E\left( {\ln \left( {\tfrac{1}{{M{{(\pi {\sigma ^2})}^M}}}\sum\nolimits_m {{e^{ - {{\left\| {{\mathbf{w}} + {\Delta _{k,m}}} \right\|}^2}/{\sigma ^2}}}} } \right)} \right)} . 
\end{align}}
By taking $h({\mathbf{y}})$ and $h({\mathbf{y}}|s)$ into (\ref{eq_mi}), we can numerically compute $I(s;{\mathbf{y}})$ as
\begin{equation}
	I(s;{\mathbf{y}}) = {\log _2}M - \tfrac{M}{\ln 2} - \tfrac{1}{M}\Sigma_k {E\left( {{{\log }_2}\Sigma_m {{e^{ - {{\left\| {{\mathbf{w}} + {\Delta _{k,m}}} \right\|}^2}/{\sigma ^2}}}} } \right)}. 
\end{equation}

\vspace{-2mm}
\section{Delay-Doppler-Angle Estimation Scheme at JCR Transceiver}\label{sec_ParaEst}

This section presents the parameter estimation scheme, serving as the foundation for tracking and data demodulation presented in Section~\ref{sec_JointTrackDataDemod}.
\vspace{-2mm}
\subsection{Problem Formulation}\label{S5.1}

  The delay-Doppler-angle estimation problem can be formulated as\vspace{-1mm}
{\small\begin{align}\label{eq_problemDIR} 
	&\arg \min\limits_{\{ \hat{r}_l^{(\cdot)},\hat{v}_l^{(\cdot)},\hat{\theta}_{\text{rx},l}^{(\cdot)},\hat{\varphi}_{\text{rx},l}^{(\cdot)},\hat{\theta}_{\text{tx},l}^{(\cdot)},\hat{\varphi}_{\text{tx},l}^{(\cdot)}\} } \!\! \sum\limits_{n_f = 0}^{N_f - 1} \sum\limits_{n_c = 0}^{N_c - 1} \sum\limits_{n_{\text{rx}} = 0}^{N_{\text{rx}} - 1} \bigg\| {\cal Y}^{({\cdot})}[n_f,n_c,n_{\text{rx}}] \nonumber\\
	&- \sum\limits_{l = 1}^{\hat{L}^{(\cdot)}} \hat{\beta}_l^{(\cdot)} \sum\limits_{n_{\text{tx}} = 0}^{N_{\text{tx}} - 1} {\bf{b}}_l^{({N_f})}[n_f]\left(\hat{f}_{r,l}^{\text{(Tl)}}\right){\bf{b}}^{({N_c})}\left[ n_c \right] \left({N_c}{n_{\text{tx}}}/{N_{\text{tx}}}\right)\nonumber\\
	&\times{\bf{b}}_l^{({N_c})}[{n_c}]\left({{\hat f}_{v,l}^{\text{(Tl)}}}\right)
	\bar{\bf{P}}_l^{(\cdot)}[{n_{\text{rx}}},{n_{\text{tx}}}]\left({{\hat \theta }_{\text{tx},l}^{(\cdot)}},{{\hat \varphi }_{\text{tx},l}^{(\cdot)}},{{\hat \theta }_{\text{rx},l}^{(\cdot)}},{{\hat \varphi }_{\text{rx},l}^{(\cdot)}}\right)\bigg\|_2^2 ,\nonumber\\
	&\text{s.t. }  \hat{r}_l^{(\cdot)} = c{{\hat f}_{r,l}^{\text{(Tl)}}}/(\xi{N_f}{T_s}\alpha ),\; {{\hat v}_l^{(\cdot)}} = c{{\hat f}_{v,l}^{\text{(Tl)}}}/(\xi{N_c}{{\bar T}_c}{f_c}), \nonumber\\
	 &\hspace{4mm}{{\hat f}_{r,l}^{\text{(Tl)}}} = \hat f_{r,l}^{( \cdot )} + \hat f_r^{({\text{D}})},\;{{\hat f}_{v,l}^{\text{(Tl)}}} = \hat f_{v,l}^{( \cdot )} + \hat f_v^{({\text{D}})},
\end{align}}
where the superscript ``($\cdot$)'' is either ``(A)'' or ``(P)'' depending on whether it represents the AV's received signal or the PV's received signal,
  $\xi\! =\! 2$ or $\xi\! =\! 1$ depending on whether it is the received signal of the AV or the PV, and the index $l$ in the case ``$(\text{A})$'' indicates the $l$-th target and there are a total of $\hat{L}^{({\text{A}})}$ targets, while the index $l$ in the case ``$(\text{P})$'' indicates the $l$-th path and there are a total of $\hat{L}^{({\text{P}})}$ paths.
 Also $\hat{\beta}_l^{(\text{A})}$, $\hat{r}_l^{(\text{A})}$, $\hat{v}_l^{(\text{A})}$, $\hat{\theta}_{\text{rx},l}^{(\text{A})}$, $\hat{\varphi}_{\text{rx},l}^{(\text{A})}$, $\hat{\theta}_{\text{tx},l}^{(\text{A})}$ and $\hat{\varphi}_{\text{tx},l}^{(\text{A})}$ are the estimates of complex amplitude, distance, radial-velocity, AoA, EoA, AoD and EoD of the $l$-th target of the AV, respectively, while $\hat{\beta}_l^{(\text{P})}$, $\hat{r}_l^{(\text{P})}$, $\hat{v}_l^{(\text{P})}$, $\hat{\theta}_{\text{rx},l}^{(\text{P})}$, $\hat{\varphi}_{\text{rx},l}^{(\text{P})}$, $\hat{\theta}_{\text{tx},l}^{(\text{P})}$ and $\hat{\varphi}_{\text{tx},l}^{(\text{P})}$ are the estimates of complex amplitude, distance, radial-velocity, AoA, EoA, AoD, and EoD of the $l$-th path of the PV, respectively.
  Additionally, ${\hat f}_{r,l}^{\text{(Tl)}}$ is the estimated normalized total delay (distance) frequency which comprises two parts, $\hat{f}_{r,l}^{( \cdot )}$ and $\hat{f}_{r}^{({\text{D}})}$, where $\hat{f}_{r,l}^{( \cdot )}$ is the part that comes from the true target (path) parameters, while $\hat{f}_{r}^{({\text{D}})}$ is the part that comes from the modulated data.
  The same is true for the estimated normalized total Doppler (radial-velocity) ${\hat f}_{v,l}^{\text{(Tl)}}$.
  Furthermore, in (\ref{eq_problemDIR}), ${\bf{b}}^{(N)}[n](f)\! =\! e^{\textsf{j}2\pi n f/N}$, ${\bf{b}}_{l}^{(N)}[n](f_l)\! =\!e^{\textsf{j}2\pi nf_l/N}$, $\bar{\bf{P}}_l^{(\cdot)} [{n_{\text{rx}}},{n_{\text{tx}}}]\left(\hat\theta _{\text{tx},l}^{(\cdot)},\hat\varphi _{\text{tx},l}^{(\cdot)},\hat\theta _{\text{rx},l}^{( \cdot )},\hat\varphi _{\text{rx},l}^{( \cdot )}\right)\! =\! {e^{\textsf{j}2\pi {f_c}/c\left( {{{\left( {\hat{\bf{d}}_{\text{tx},l}^{(\cdot)}} \right)}^{\text T}}{{\bf{p}}_{\text{tx}}}[{n_{\text{tx}}}] + {{\left( {\hat{\bf{d}}_{\text{rx},l}^{( \cdot )}} \right)}^{\text T}}{{\bf{p}}_{\text{rx}}}[{n_{\text{rx}}}]} \right)}}$, $\mathbf{\hat d}_{\text{tx},l}^{(\cdot)}\! =\! \big[\cos \hat\varphi_{\text{tx},l}^{(\cdot)} \sin \hat\theta_{\text{tx},l}^{(\cdot)}, $
  $\cos \hat\varphi_{\text{tx},l}^{(\cdot)} \cos \hat\theta_{\text{tx},l}^{(\cdot)}, \sin \hat\varphi_{\text{tx},l}^{(\cdot)}\big]^{\text T}$, and $\mathbf{\hat d}_{\text{rx},l}^{(\cdot)}\! =\! \big[\cos \hat\varphi_{\text{rx},l}^{(\cdot)} \sin \hat\theta_{\text{rx},l}^{(\cdot)} , $
  $\cos \hat\varphi_{\text{rx},l}^{(\cdot)} \cos \hat\theta_{\text{rx},l}^{(\cdot)} ,\ \sin \hat\varphi_{\text{rx},l}^{(\cdot)}\big]^{\text T}$.

  Since the detection is performed on the RDM, the detected position can be directly used to derive the corresponding delay and Doppler.
  Therefore, the problem in (\ref{eq_problemDIR}) can be efficiently solved by first addressing the detection sub-problem, followed by applying some angle estimation algorithms.
  In light of this, cell averaging-CFAR is used to obtain the delay and Doppler in the RDM. However, spectrum leakage can lead to multiple detected positions for a single target (path).
  To address this issue, we propose the ``beacon frame-aided 4D-parameter estimation scheme", which leverages spectrum leakage rather than being adversely affected by it.
  Note that the PV can only obtain $\hat{f}_{r,l}^{({\text{\text{Tl}}})}$ and $ \hat{f}_{v,l}^{({\text{Tl}})}$ in this scheme.
  $\hat{f}_r^{({\text{D}})}$ and $\hat{f}_v^{({\text{D}})}$ are obtained by the PV using the scheme of Section~\ref{sec_JointTrackDataDemod}, since tracking process can provide prior information of $\hat{f}_{r,l}^{({\text{P}})}$ and $\hat{f}_{v,l}^{({\text{P}})}$.
\begin{algorithm}[!t]
	\label{alg_ParaEst} 
	\begin{scriptsize}
		\caption{Proposed Beacon Frame-Aided 4D-Parameter Estimation Scheme}
		\LinesNumbered
		\KwIn{Received signal after 2D DFT ${\mathcal{\underline Y}}^{({\cdot})}\in \mathbb{C}^{N_f\times N_c\times N_{\text{rx}}}$ and ${\mathcal{\tilde {\underline Y}}}^{({\cdot})}\in \mathbb{C}^{N_f\times N_c\times N_{\text{rx}}}$, RDM after CFAR detection ${{\bf{Y}}}_{{\text{RDM,D}}}^{({\cdot})}\in \mathbb{C}^{N_f\times N_c}$ and ${\tilde{\bf{Y}}}_{{\text{RDM,D}}}^{({\cdot})}\in \mathbb{C}^{N_f\times N_c}$, Thresholds for determining whether it is the same target (path) $\varpi_r$ and $\varpi_v$; }
		\KwOut{4D parameters for $l$-th target (path) $\big\{ {\bar{\hat{r}}}_l^{(\cdot)},{\bar{\hat{v}}}_l^{(\cdot)},{\bar{\hat{\theta}}}_{\text{rx},l}^{(\cdot)},{\bar{\hat{\varphi}}}_{\text{rx},l}^{(\cdot)},{\bar{\hat{\theta}}}_{\text{tx},l}^{(\cdot)},{\bar{\hat{\varphi}}}_{\text{tx},l}^{(\cdot)}\big\}$, $l\in \{1,\cdots,\hat{L}^{(\cdot)}\}$; }
		\For{$i_f=\{ 0, 1, \cdots, N_f-1\}$}{
			\For{$i_c=\{ 0, 1, \cdots,N_c-1 \}$}{		
				Obtain distance and velocity associated with target (path) on ($i_f$, $i_c$)-th grid in RDM as (\ref{eq_SL2T_CFAR_second_start})--(\ref{eq_SL2T_CFAR_second_end})\;
				Obtain DoA and DoD associated with target (path) on ($i_f$, $i_c$)-th grid in the RDM as (\ref{eq_ArrayBeforeRearrange})--(\ref{eq_SL2T_CFAR_third_end})\;}}
		Obtain final 4D parameters of each target (path) by averaging corresponding target (path) lists as (\ref{eq_SL2T_CFAR_final}).
	\end{scriptsize}
\end{algorithm}\vspace{-5mm}
\vspace{-2mm}
\subsection{Beacon Frame-Aided 4D-Parameter Estimation}\label{sec_BeaconFrameAidedParaEst}

  The complete process using ${\tilde{\bf{Y}}}_{{\text{RDM,D}}}^{(\cdot)}$, ${{\bf{Y}}}_{{\text{RDM,D}}}^{({\cdot})}$, ${\mathcal{\tilde{\underline{Y}}}}^{(\cdot)}$ and ${\mathcal{\underline Y}}^{(\cdot)}$ to obtain the target (path) list with 4D parameters is summarized as the beacon frame-aided 4D-parameter estimation scheme in Algorithm~\ref{alg_ParaEst}.
  ${\tilde{\bf{Y}}}_{{\text{RDM,D}}}^{({\cdot})}$ and ${\mathcal{\tilde{\underline{Y}}}}^{({\cdot})}$ are obtained from the beacon frame, while ${{\bf{Y}}}_{{\text{RDM,D}}}^{({\cdot})}$ and  ${\mathcal{\underline Y}}^{({\cdot})}$ are obtained from the DDM frame, which follows the beacon frame immediately. 

	Due to spectrum leakage, each target (path) estimated from the RDM has multiple similar parameter values. Therefore, each 4D parameter for each target (path) is stored in an individual list. This process turns waste of spectrum leakage into treasure, since we can obtain more precise estimates through averaging all the values in the list. Since there are six different 4D parameters, each target (path) has six lists. Algorithm~\ref{alg_ParaEst} maintains $\hat L^{(\cdot)}\times 6$ lists for $\hat{L}^{(\cdot)}$ targets (paths).
	For each target (path), the six lists of 4D parameters, i.e., $\hat{\mathbf{r}}_l^{(\cdot)}$, $\hat{\mathbf{v}}_l^{(\cdot)}$, $\hat{\bm{\theta}}_{\text{tx},l}^{(\cdot)}$, $\hat{\bm{\varphi}}_{\text{tx},l}^{(\cdot)}$, $\hat{\bm{\theta}}_{\text{rx},l}$ and $\hat{\bm{\varphi}}_{\text{rx},l}^{(\cdot)}$, $l\! \in\! \{1,\cdots,\hat{L}^{(\cdot)}\}$, contain the same number of elements. However, due to varying levels of spectrum leakage, the number of elements in these lists differs among different targets (paths).
	Before performing CFAR detection on the RDM, $\hat{L}^{(\cdot)}\! =\! 0$. Then $\hat{L}^{(\cdot)}$ is continuously updated in the subsequent steps.

  For each $i_f\! =\! \{ 0, 1, \cdots, N_f-1\}$ and $i_c\! =\! \{ 0, 1, \cdots,N_c-1 \}$ in ${\tilde{\bf{Y}}}_{{\text{RDM,D}}}^{(\cdot)}$ and ${\bf{Y}}_{{\text{RDM,D}}}^{({\cdot})}$, we estimate the distance and the velocity of the corresponding target (path) first. If\vspace{-1mm}
{\footnotesize\begin{align}\label{eq_SL2T_CFAR_second_start}	
	 {\bf{Y}}_{{\text{RDM,D}}}^{( \cdot )} \left[{i_f},({i_c}:{N_c}/{N_{\text{tx}}}:{i_c} + ({N_{\text{tx}}} - 1){N_c}/{N_{\text{tx}}}) \mod {N_c}\right]= {\bf{1}}_{{N_{\text{tx}}}}^{\text T}	
\end{align}}
and\vspace{-1mm}
{\footnotesize\begin{align}	\label{eqACFAR1} 
	{\text{Avg}}\big(  \tilde{\mathbf{Y}}_{{\text{RDM,D}}}^{( \cdot )}\big[({i_f} - 1:{i_f} + 1)\hspace{-2mm} \mod \hspace{-1mm}{N_f},({i_c} - 1:{i_c} + 1)\hspace{-2mm} \mod\hspace{-1mm} {N_c}\big] \big)\ne 0
\end{align}}\vspace{-1mm}
are satisfied, the $(i_f, i_c)$-th grid in ${\bf{Y}}_{{\text{RDM,D}}}^{( \cdot )}$ is considered having a target (path).
  If $\hat{L}^{(\cdot)}\! =\! 0$, we assume the presence of the first target (path) on the $(i_f, i_c)$-th grid and create parameters as
{\small\begin{equation}\label{eqACFAR2} 
	l=1;\;\hat L^{(\cdot)} = 1;\;{ {\bf{\hat r}}_{l}}^{(\cdot)} = \left[r_{\text{res}}^{( \cdot )}{i_f}\right];\;{ {\bf{\hat v}}_{l}}^{(\cdot)} = \left[v_{\text{res}}^{( \cdot )}({i_c} - {N_c}/2)\right].
\end{equation}}
If $\hat{L}^{(\cdot)}\! \ne\! 0$, we search for the $l$-th target (path) in the existing target (path) lists that has the closest distance to $i_f$ and $i_c$ according to 
{\footnotesize\begin{align}\label{eqACFAR3} 
	\arg \min\limits_l \left|{\text{Avg}}({ {\bf{\hat r}}_l^{(\cdot)}})/r_{\text{res}}^{( \cdot )} - {i_f}\right|\! +\! \left|{\text{Avg}}({{\bf{\hat v}}_l^{(\cdot)}})/v_{\text{res}}^{( \cdot )}\! -\! ({i_c}\! -\! {N_c}/2)\right|,
\end{align}}
where $r_{\text{res}}^{( \cdot )}$ and $v_{\text{res}}^{( \cdot )}$ are the distance and velocity resolutions given in Table~\ref{table_SensingBound}.
  If both conditions $|{\text{Avg}}({ {\bf{\hat r}}_l^{(\cdot)}})/r_{\text{res}}^{( \cdot )} - {i_f}| \le {\varpi _r}$ and $|{\text{Avg}}({ {\bf{\hat v}}_l^{(\cdot)}})/v_{\text{res}}^{( \cdot )}\! -\! {i_c}\! -\! \frac{{{N_c}}}{2}| \le {\varpi _v}$ are met, where ${\varpi _r}$ and ${\varpi _v}$ are predefined thresholds, we include the $(i_f, i_c)$-th grid into the previous $l$-th target (path) lists as
\begin{equation}\label{eqACFAR3} 
	{\bf{\hat{r}}}_l^{(\cdot)} = [{ {\bf{\hat r}}_l^{(\cdot)}},\;r_{\text{res}}^{( \cdot )}{i_f}];\ { {\bf{\hat v}}_l^{(\cdot)}} = [{ {\bf{\hat v}}_l^{(\cdot)}},v_{\text{res}}^{( \cdot )}({i_c} - {N_c}/2)].
\end{equation}
Otherwise, we assume the presence of a new target (path) on the $(i_f, i_c)$-th grid and we create parameters as
\begin{align}\label{eq_SL2T_CFAR_second_end} 
	&l=\hat L^{(\cdot)} + 1;\;\hat L^{(\cdot)} = \hat L^{(\cdot)} + 1;\;\nonumber\\
	&{ {\bf{\hat r}}_{l}}^{(\cdot)} = \left[r_{\text{res}}^{( \cdot )}{i_f}\right];\;{ {\bf{\hat v}}_{l}}^{(\cdot)} = \left[v_{\text{res}}^{( \cdot )}({i_c} - {N_c}/2)\right].
\end{align}

  Second, we proceed to estimate angles of the $l$-th target (path) whose distance and velocity have already been estimated in the preceding process.
  We are now detecting the $u$-th detection position of the $l$-th target (path), where $u\! =\! \text{len}\left({{\bf{ {\hat r}}}_{l}^{(\cdot)}}\right)\! =\! \text{len}\left({{\bf{ {\hat v}}}_{l}^{(\cdot)}}\right)$.
  The complex amplitudes of signals transmitted by different antennas used to estimate angles on the same receive antenna can be extracted as
\begin{align}\label{eq_ArrayBeforeRearrange} 
	 {\tilde{\bf{P}}}[n_{\text{rx}},:]\! &=\! {\mathcal{\underline Y}}^{({\cdot})}\big[ {\bf{\hat r}}_{l}^{(\cdot)} [u]/r_{\text{res}}^{( \cdot )},\big( {\bf{\hat{v}}}_{l}^{(\cdot)} [u]/v_{\text{res}}^{( \cdot )} \nonumber\\
	&+ (0:{N_c}/{N_{\text{tx}}}:{{N}_c}\! -\! {N_c}/{N_{\text{tx}}}) \big)\!\!\! \mod N_c, n_{\text{rx}}\big],	\!
\end{align}		
where $\tilde{\mathbf{P}}\in\mathbb{C}^{N_{\text{rx},a}N_{\text{rx},e}\times N_{\text{tx},a}N_{\text{tx},e}}$.

  If ``$(\cdot)$'' is ``$\text{(A)}$'', we can extend the antenna aperture, since AoA equals AoD and EoA equals EoD.
$\tilde {\bf{P}}$'s columns represent the indices of the transmit antennas, while $\tilde {\bf{P}}$'s rows represent the indices of the receive antennas. The elements in $\tilde{\bf{P}}$ are then reordered to facilitate angle estimation as
\begin{align}\label{eq_ArrayAfterRearrange} 
	& {\bf{\bar P}}[ \left\lfloor {n_{\text{tx}}}/{N_{\text{tx},a}}\right\rfloor N_{\text{rx},e}\! +\! \left\lfloor {n_{\text{rx}}}/{N_{\text{rx},a}} \right\rfloor, ({n_{\text{tx}}}\!\!\! \mod{N_{\text{tx},a}}){N_{\text{rx},a}}\! +\nonumber\\
	&\! \left({n_{\text{rx}}}\!\!\! \mod {N_{\text{rx},a}\right)} ] \Leftarrow \tilde{\bf{ P}}[{n_{\text{rx}}},{n_{\text{tx}}}],
\end{align}
where $\mathbf{\bar P}\in\mathbb{C}^{N_{\text{tx},e}N_{\text{rx},e}\times N_{\text{tx},a}N_{\text{rx},a}}$\footnote{By using the AIC and MDL criteria\cite{ref_AIC_MDL}, we can determine the number of different angles with the same distance and velocity in $\mathbf{\bar P}$. For simplicity, we set the number of angles with the same distance and velocity as one in this paper.}.
  $\mathbf{\bar P}$ is the array response of the extended array formed by transmit antennas and receive antennas, resulting in the improvement of angle estimation performance.

  If ``$(\cdot)$'' is ``$\text{(P)}$'', we cannot extend the antenna aperture since DoA no longer equals DoD.
However, we can also utilize the dimension brought by the transmit antennas, since the columns or rows of $\tilde{\mathbf{P}}$ can be considered as multiple ``snapshots" in array signal processing. Therefore, subspace super-resolution algorithms can be used to estimate the DoA and DoD.

  However, regardless whether ``$(\cdot)$'' is ``(A)'' or ``(P)'', we choose to use DFT with redundant-dictionary to estimate the DoA and DoD with lower complexity \cite{ref_LZR}.
  While this approach yields on-grid estimates within each frame, the tracking process described in Section~\ref{sec_JointTrackDataDemod} enables off-grid estimation, which is validated by our simulations in Section~\ref{Sec_Simulation}.
  We define the estimated AoA, EoA, AoD, and EoD as ${\hat{\theta}}_{\text{rx},i_f,i_c}^{(\cdot)}$, ${\hat{\varphi}}_{\text{rx},i_f,i_c}^{(\cdot)}$, ${\hat{\theta}}_{\text{tx},i_f,i_c}^{(\cdot)}$ and ${\hat{\varphi}}_{\text{tx},i_f,i_c}^{(\cdot)}$, respectively. We include them into the $l$-th target (path) lists as
\begin{align}\label{eq_SL2T_CFAR_third_end} 
\left\{\begin{array}{ll}
	{\bm{\hat{\theta}}}_{\text{tx},l}^{(\cdot)} = \big[{\bm{\hat{\theta}}}_{\text{tx},l}^{(\cdot)}, {\hat{\theta}}_{\text{tx},i_f,i_c}^{(\cdot)}\big]; & {\bm{\hat{\varphi}}}_{\text{tx},l}^{(\cdot)} = \big[{\bm{\hat{\varphi}}}_{\text{tx},l}^{(\cdot)}, {\hat{\varphi}}_{\text{tx},i_f,i_c}^{(\cdot)}\big]; \\
	{\bm{\hat{\theta}}}_{\text{rx},l}^{(\cdot)} = \big[{\bm{\hat{\theta}}}_{\text{rx},l}^{(\cdot)}, {\hat{\theta}}_{\text{rx},i_f,i_c}^{(\cdot)}\big]; & {\bm{\hat{\varphi}}}_{\text{rx},l}^{(\cdot)} = \big[{\bm{\hat{\varphi}}}_{\text{rx},l}^{(\cdot)}, {\hat{\varphi}}_{\text{rx},i_f,i_c}^{(\cdot)}\big].
\end{array}\right.
\end{align}

  Finally, after detecting the $N_f\! \times\! N_c$ grids in ${\tilde{\bf{Y}}}_{{\text{RDM,D}}}^{(\cdot)}$ and ${{\bf{Y}}}_{{\text{RDM,D}}}^{({\cdot})}$ through the above procedure, the final 4D parameters for the $l$-th target (path) are obtained by averaging the corresponding lists as
\begin{footnotesize}
\begin{align}\label{eq_SL2T_CFAR_final} 
\left\{\begin{array}{lll}
	{\bar{\hat{r}}}_l^{(\cdot)} = {\text{Avg}}\big({\bm{\hat{r}}}_l^{(\cdot)}\big), & {\bar{\hat{v}}}_l^{(\cdot)} = {\text{Avg}}\big({\bm{\hat{v}}}_l^{(\cdot)}\big), & {\bar{\hat{\theta}}}_{\text{tx},l}^{(\cdot)} = {\text{Avg}}\big({\bm{\hat{\theta}}}_{\text{tx},l}^{(\cdot)}\big), \\
	{\bar{\hat{\varphi}}}_{\text{tx},l}^{(\cdot)} = {\text{Avg}}\big({\bm{\hat{\varphi}}}_{\text{tx},l}^{(\cdot)}\big), & {\bar{\hat{\theta}}}_{\text{rx},l}^{(\cdot)} = {\text{Avg}}\big({\bm{\hat{\theta}}}_{\text{rx},l}^{(\cdot)}\big), & {\bar{\hat{\varphi}}}_{\text{rx},l}^{(\cdot)} = {\text{Avg}}\big({\bm{\hat{\varphi}}}_{\text{rx},l}^{(\cdot)}\big).
\end{array}\right.
\end{align}
\end{footnotesize}
\begin{remark}\label{R3}
	Although the aforementioned process requires both the beacon frame and the DDM frame, the role of the beacon frame is to unambiguously determine velocities in the RDM.
	Therefore, upon obtaining 4D parameters in the initial channel sensing stage, only the DDM frame is needed in the subsequent process.
	This is because the 4D parameters estimated in the last DDM frame can be used to unambiguously determine velocities in the RDM of the current frame, with the last DDM frame serving as the role of the beacon frame.
	\hfill\qedsymbol
\end{remark}

\vspace{-4mm}
\subsection{Convergence and Complexity Analysis}
	Algorithm~\ref{alg_ParaEst} performs detection on the RDM. The two `for' loops are designed to systematically process each detection point on the RDM. Each step within the loops is a deterministic process without any iteration. Therefore, the algorithm is guaranteed to complete and thus converges.
	
The overall computational complexity of Algorithm~\ref{alg_ParaEst} can be analyzed by breaking it down into its primary components. First, the generation of the RDM from the received signal, as described in (\ref{eq_2D_DFT}), requires performing a 2D-DFT for each of the $N_{\text{rx}}$ receive antennas, resulting in a complexity of $\mathcal{O}(N_{\text{rx}}N_fN_c(\log N_f + \log N_c))$. Subsequently, CFAR detection is applied to the RDM, which has a linear complexity of $\mathcal{O}(N_fN_c)$ with respect to the RDM size. The following step involves iterating through each grid cell of the RDM to perform target/path association ((\ref{eq_SL2T_CFAR_second_start})--(\ref{eqACFAR1})), which also scales linearly with the RDM size, contributing $\mathcal{O}(N_fN_c)$ to the complexity. Finally, for the $\hat{L}^{(\cdot)}$ targets/paths identified, angle estimation is performed using a DFT-based search, leading to a complexity of $\mathcal{O}(\hat{L}^{(\cdot)} \cdot N_{\text{tx}}N_{\text{rx}}(G_\theta + G_\varphi))$, where $G_{\theta}$ and $G_{\varphi}$ are the number of search grids for azimuth and elevation, respectively. Consequently, the total complexity is the summation of these stages: $\mathcal{O}(N_{\text{rx}}N_fN_c(\log N_f + \log N_c)) + \mathcal{O}(N_fN_c) + \mathcal{O}(\hat{L}^{(\cdot)} \cdot N_{\text{tx}}N_{\text{rx}}(G_\theta + G_\varphi))$.

\vspace{-3mm}
\section{Joint Tracking and Data Demodulation}\label{sec_JointTrackDataDemod}

\subsection{Proposed Extended Kalman Filter-Based 5D Parameter Estimation for Active Vehicle}\label{sec_EKF4RT}

  The goal of JCR in the AV side is to track the interested targets by the chirp waveform carrying modulated data.
4D parameters can be obtained with at most two frames by using Algorithm~\ref{alg_ParaEst}.
  As a form of information fusion, tracking, which is the integration of the current observation with historical prior information, can obtain more accurate results than those obtained independently from two sources \cite{ref_StatisticalSP}.
  Therefore, we adopt tracking, aiming to obtain the orientation, tangential-velocity and more accurate estimation of the other 4D parameters. The combination of tangential-velocity with 4D parameters, including distance, radial-velocity, azimuth angle and elevation angle, constitutes 5D parameters.
  Compared to \cite{ref_TAES_HorizontalVel}, our approach leverages time series information to estimate the tangential-velocity, thus alleviating additional hardware requirements.

\begin{figure}[!t]
	\centering
	\includegraphics[width=2.5in]{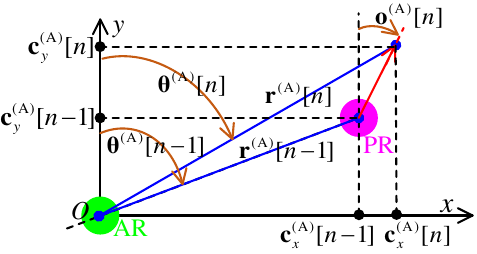}
	\vspace{-5mm}
	\caption{Geometric relationship and key parameters of the relative motion of the AV and the PV from the perspective of AV. For simplicity and clarity, other targets are not depicted.}
	\label{fig_geometry} 
	\vspace{-8mm}
\end{figure}

  All targets are tracked by the AV. The following tracking example focuses on one of the targets for illustration, but the description in this subsection applies to all targets and tracking all targets are considered in the simulations. For simplicity, we omit the subscript ``$l$''. 
  The tracking problem to be solved by the AV can be formulated as a Bayes minimum mean square error problem as
\begin{align}\label{eq_BMMSE} 
	\arg \min\nolimits_{{\bf{\hat{S}}}^{\text{(A)}}[i,n]} E\left[\big({\bf{S}}^{\text{(A)}}[i,n] - {\bf{\hat{S}}}^{\text{(A)}}[i,n]\big)^2\right],
\end{align}
where ${\hat{\bf{S}}}^{\text{(A)}}[i,n]$ is the estimated $i$-th state variable of one target in the $n$-th frame, while ${{\bf{S}}_i^{\text{(A)}}[:,n]}$ is the true one\footnote{Although $s(t)$ is used to represent the chirp signal, we use $\mathbf{S}$ to represent state variable to avoid too many symbol definitions, where ambiguities can be resolved through context. This applies to other symbol definitions as well.}.
  We define the estimated state vector in the $n$-th frame as
\begin{equation}\label{eq_StaVec} 
	\mathbf{\hat S}^{\text{(A)}}[:,n] = \left[ \mathbf{\hat c}_x^{\text{(A)}}[n],\ \mathbf{\hat c}_y^{\text{(A)}}[n],\ \mathbf{\hat v}_x^{\text{(A)}}[n],\ \mathbf{\hat v}_y^{\text{(A)}}[n]\right]^{\text T} ,
\end{equation}
where $\mathbf{\hat c}_x^{\text{(A)}}[n]$, $\mathbf{\hat c}_y^{\text{(A)}}[n]$, $\mathbf{\hat v}_x^{\text{(A)}}[n]$ and $\mathbf{\hat v}_y^{\text{(A)}}[n]$ are the estimates of the target's distance in the $x$-direction, distance in the $y$-direction, velocity in the $x$-direction and velocity in the $y$-direction, respectively.
  As illustrated in Fig.~\ref{fig_geometry}, the origin of the reference frame is the AV itself, the forward direction of the AV is the positive direction of the $y$-axis, and the positive $x$-axis is perpendicular to the $y$-axis. According to \cite{ref_StatisticalSP}, EKF can be used to solve the aforementioned problem.

  Without loss of generality and to simplify the problem, we only track the azimuth angle and assume that all targets are moving in the horizontal plane.
  Since the modulated data is known by the AV, AV can remove the pre-modulated data from the echo, obtaining a return signal only ``modulated'' by the target's delay, velocity, and angle as
{
	\begin{equation}\label{eq_RmDataRT} 
	{{\mathcal{Y}}^{({\text{A}})}}[{n_f},{n_c},{n_{\text{rx}}}] \Leftarrow \frac{{{{\mathcal{Y}}^{({\text{A}})}}[{n_f},{n_c},{n_{\text{rx}}}]}}{{\beta _{}^{({\text{D}})}{e^{\textsf{j}2\pi \frac{{{n_f}}}{{{N_f}}}f_r^{({\text{D}})} + \textsf{j}2\pi \frac{{{n_c}}}{{{N_c}}}f_v^{({\text{D}})}}}}},
\end{equation}}
where ${{\mathcal{Y}}^{({\text{A}})}}[{n_f},{n_c},{n_{\text{rx}}}]$ is replaced by ${\tilde{\mathcal{Y}}^{({\text{A}})}}[{n_f},{n_c},{n_{\text{rx}}}]$ if the frame is a beacon frame. Then, Algorithm~\ref{alg_ParaEst} is used to estimate the 4D parameters of targets.
  To minimize the Bayes minimum mean square error, we adopt the sequential linear minimum mean square error estimation. For illustrative purpose, we consider the tracking process of one target from the target lists.
  The observation vector $\mathbf{\hat X}^{\text{(A)}}[:,n]$ in the $n$-th frame is the partial outputs of Algorithm~\ref{alg_ParaEst} given by
\begin{equation}\label{eq_ObsVec} 
	\mathbf{\hat X}^{\text{(A)}}[:,n]=\left[\mathbf{\hat r}^{\text{(A)}}[n],\ \mathbf{\hat v}_r^{\text{(A)}}[n],\ \boldsymbol{\hat \theta^{\text{(A)}}}[n]\right]^{\text T},
\end{equation}
where $\mathbf{\hat r}^{\text{(A)}}[n], \mathbf{\hat v}_r^{\text{(A)}}[n], \text{and } \bm{\hat \theta^{\text{(A)}}}[n]$ are the estimates of the target's distance, radial-velocity and AoA, respectively.

  The state equation and observation equation can be formulated as
\begin{equation}\label{eq_StateEqObservationEq} 
\left\{ \begin{array}{cl}
  \mathbf{\hat{S}}^{\text{(A)}}[:,n] & = \mathbf{F}^{\text{(A)}} \mathbf{\hat{S}}^{\text{(A)}}[:,n-N_s^{\text{(A)}}] + \mathbf{u}^{\text{(A)}}, \\
	\mathbf{\hat{X}}^{\text{(A)}}[:,n] & = \mathbf{h}^{\text{(A)}}\big(\mathbf{\hat{S}}^{\text{(A)}}[:,n]\big) +\mathbf{w}^{\text{(A)}} ,
\end{array} \right.
\end{equation}
where $\mathbf{u}^{\text{(A)}}\! \sim\! \mathcal{N}(\mathbf{0}, \mathbf{C}_u^{\text{(A)}})$ is the state perturbation with the diagonal covariance matrix $\mathbf{C}_u^{\text{(A)}}$ and $\mathbf{w}^{\text{(A)}}\!\sim\! \mathcal{N}(\mathbf{0}, \mathbf{C}_w^{\text{(A)}})$ is the observation noise with  the diagonal covariance matrix $\mathbf{C}_w^{\text{(A)}}$, while $N_s^{\text{(A)}}$ represents that fusion is performed every $N_s^{\text{(A)}}$ frames by the AV.
  Note that for the sake of simplification, we assume that errors across different states (or observations) are uncorrelated, and their mean values are zero. Also since the target to be tracked does not undergo significant changes between consecutive frames, we conduct tracking once every $N_s^{\text{(A)}}$ frames, rather than tracking in each frame to reduce computation complexity.
 In (\ref{eq_StateEqObservationEq}), $\mathbf{h}^{\text{(A)}}(\cdot)$ is the observation function, whose input is the estimated state of the $n$-th frame and whose output is the observation of the $n$-th frame, and $\mathbf{F}^{\text{(A)}}$ is the state transition matrix given by
{\begin{small}\begin{align}\label{eqSTM} 
	{\bf{F}^{\text{(A)}}} = \left[ \begin{array}{*{20}{c}}
			1&0&N_s^{\text{(A)}}\bar{T}_c&0\\
			0&1&0&N_s^{\text{(A)}}\bar{T}_c\\
			0&0&1&0\\
			0&0&0&1
	\end{array} \right] .
\end{align}\end{small}}
Furthermore, the output-input relationship of the observation function $\mathbf{h}^{\text{(A)}}(\cdot)$ is defined as
{\footnotesize\begin{equation}\label{eq_OutIn} 
\left\{ \begin{array}{cl}	
	\mathbf{\hat{c}}_x^{\text{(A)}}[n] & = \mathbf{\hat r}^{\text{(A)}}[n]\cos \boldsymbol{\hat \theta^{\text{(A)}}} [n],\\
		\mathbf{\hat c}_y^{\text{(A)}}[n] & = \mathbf{\hat r}^{\text{(A)}}[n]\sin \boldsymbol{\hat \theta^{\text{(A)}}} [n],\\
		\mathbf{\hat v}_x^{\text{(A)}}[n] & = \mathbf{\hat v}_r^{\text{(A)}}[n]\sin \mathbf{\hat o}^{\text{(A)}}[n]/ \cos\left(\boldsymbol{\hat \theta^{\text{(A)}}}[n]-\mathbf{\hat o}^{\text{(A)}}[n]\right),\\
		\mathbf{\hat v}_y^{\text{(A)}}[n] & = \mathbf{\hat v}_r^{\text{(A)}}[n]\cos \mathbf{\hat o}^{\text{(A)}}[n]/ \cos\left(\boldsymbol{\hat \theta^{\text{(A)}}}[n]-\mathbf{\hat o}^{\text{(A)}}[n]\right) ,
\end{array} \right.
\end{equation}}
where $\mathbf{\hat{o}}^{\text{(A)}}[n]$ is the target's orientation given by
\begin{align}\label{eqTO} 
	\mathbf{\hat{o}}^{\text{(A)}}[n] =& {\arcsin}\left({\Delta x}/({\left(\Delta x\right)^2}+{\left(\Delta y\right)^2})^{0.5}\right)	
\end{align}
in which $\Delta x\! =\! {\mathbf{\hat c}_x^{\text{(A)}}[n]-\mathbf{\hat c}_x^{\text{(A)}}[n-N_o]}$, $\Delta y\! =\! \mathbf{\hat c}_y^{\text{(A)}}[n]-\mathbf{\hat c}_y^{\text{(A)}}[n-N_o]$, and $N_o$ is the number of frames used as an interval for orientation estimation.

We can obtain the input-output relationship of the observation function $\mathbf{h}^{\text{(A)}}(\cdot)$ through (\ref{eq_OutIn}) as
{\footnotesize\begin{align}\label{eq_InOut}	
\left\{\begin{array}{ll}
	\mathbf{\hat{r}}^{\text{(A)}}[n] & = \left(\left(\mathbf{\hat c}_x^{\text{(A)}}\right)^2[n] + \left(\mathbf{\hat c}_y^{\text{(A)}}\right)^2[n]\right)^{0.5}, \\
  \mathbf{\hat{v}}_r^{\text{(A)}}[n] & = \left\{ \begin{array}{*{20}{l}}
		\mathbf{\hat{v}}_x^{\text{(A)}}[n] \cos\left(\bm{\hat{\theta}}^{\text{(A)}} [n] - \mathbf{\hat{o}}^{\text{(A)}}[n]\right)/\sin \mathbf{\hat{o}}^{\text{(A)}}[n], \\ \hspace{36mm}\text{if}\;|\mathbf{\hat{o}}^{\text{(A)}}[n]| > \pi /4,\\
		\mathbf{\hat{v}}_y^{\text{(A)}}[n] \cos \left(\bm{\hat{\theta}}^{\text{(A)}}[n] - \mathbf{\hat{o}}^{\text{(A)}}[n]\right)/\cos \mathbf{\hat{o}}^{\text{(A)}}[n],\\ \hspace{36mm}\text{if}\;|\mathbf{\hat{o}}^{\text{(A)}}[n]| \le \pi /4,
\end{array} \right. \\
	\bm{\hat{\theta}}^{\text{(A)}} [n] & = \arcsin \left(\mathbf{\hat{c}}_x^{\text{(A)}}[n]/{ (\left(\mathbf{\hat{c}}_x^{\text{(A)}}\right)^2[n] + \left(\mathbf{\hat{c}}_y^{\text{(A)}}\right)^2[n])^{0.5} }\right),
\end{array}\right.
\end{align}}
where the condition about $\mathbf{\hat o}^{\text{(A)}}[n]$ is designed to prevent the zero denominator.
  The proposed EKF-based 5D-parameter estimation for AV is summarized in Algorithm~\ref{alg_EKF_RT}, where 
$\bm{\mu}^{\text{(A)}}_{-}[:,n]$, $\bm{\mu}^{\text{(A)}}_{+}[:,n]$, $\bm{\Sigma}^{\text{(A)}}_{-}[:,:,n]$ and $\bm{\Sigma}^{\text{(A)}}_{+}[:,:,n]$ are the predictive mean, posterior mean, predictive variance and posterior variance of the state vector $\mathbf{S}^{\text{(A)}}[:,n]$, respectively, and ${\bf{J}}^{(\text{A})}\! =\! \frac{\partial \mathbf{h}^{\text{(A)}}({\bf{S}}[:,n])}{\partial {\bf{S}}^{\text T}[:,n]}\big|_{{\bf{S}}[:,n] = \hat{\bf{S}}^{\text{(A)}}[:,n]} \! \in\! \mathbb{R}^{3\times 4}$ is the Jacobian matrix to linearize $\mathbf{h}^{\text{(A)}}({\bf{\hat{S}}}^{\text{(A)}}[:,n])$, while $\mathbf{K}^{\text{(A)}}$ is the Kalman gain used to obtain posterior mean.
	The posterior mean and variance first time used in Algorithm~\ref{alg_EKF_RT} are derived from the initial channel sensing stage. Steps~\ref{alg_EKF_RT_start}--\ref{alg_EKF_RT_end} is the standard EKF procedure \cite{ref_StatisticalSP}.
\begin{algorithm}[!t]
	\label{alg_EKF_RT} 
	\begin{scriptsize}
		\caption{Proposed Extended Kalman Filter-Based 5D Parameter Estimation for AV}
		\LinesNumbered
		\KwIn{Posterior mean vector $\bm{\mu}^{\text{(A)}}_+[:,n\! -\! N_s^{\text{(A)}}]\! \in\! \mathbb{R}^{4}$ and posterior covariance matrix $\bm{\Sigma}^{\text{(A)}}_+[:,:,n\! -\! N_s^{\text{(A)}}]\! \in\! \mathbb{R}^{4\times 4}$ of state vector $\mathbf{S}^{\text{(A)}}[:,n\! -\! N_s^{\text{(A)}}]\! \in\!  \mathbb{R}^{4}$,
			state perturbation covariance matrix $\mathbf{C}^{\text{(A)}}_u\! \in\! \mathbb{R}^{4\times 4}$, and observation noise covariance matrix $\mathbf{C}^{\text{(A)}}_w\! \in\! \mathbb{R}^{4\times 4}$; }
		\KwOut{Posterior mean vector $\bm{\mu}^{\text{(A)}}_+[:,n]\! \in\! \mathbb{R}^{4}$ and posterior covariance matrix $\bm{\Sigma}^{\text{(A)}}_+[:,:,n]\! \in\! \mathbb{R}^{4\times 4}$ of current state vector $\mathbf{S}^{\text{(A)}}[:,n]\! \in\! \mathbb{R}^{4}$; }
		Obtain received signal, ${\tilde{\mathcal{Y}}^{({\text{A}})}}$ and ${{\mathcal{Y}}^{({\text{A}})}}$, as (\ref{eq_RxSignalRT})\;
		Remove pre-modulated data of received signal as (\ref{eq_RmDataRT})\;
		Perform DFT on ${\tilde{\mathcal{Y}}^{({\text{A}})}}$ and ${{\mathcal{Y}}^{({\text{A}})}}$ to obtain ${\tilde{\mathcal{\underline{Y}}}^{({\text{A}})}}$ and ${{\mathcal{\underline{Y}}}^{({\text{A}})}}$ as (\ref{eq_2D_DFT})\;
		Obtain $\tilde{\bf{Y}}_{{\text{RDM}}}^{({\text{A}})}$ and ${\bf{Y}}_{{\text{RDM}}}^{({\text{A}})}$ as (\ref{eq_RDM})\;
		Perform CFAR detection on $\tilde{\bf{Y}}_{{\text{RDM}}}^{({\text{A}})}$ and ${\bf{Y}}_{{\text{RDM}}}^{({\text{A}})}$ to obtain ${\tilde{\bf{Y}}^{({\text{A}})}_{\text{RDM,D}}}$ and ${{\bf{Y}}^{({\text{A}})}_{\text{RDM,D}}}$\;
		Obtain observation vector $\mathbf{\hat X}^{\text{(A)}}[:,n]$ through Algorithm~\ref{alg_ParaEst}\;
		\tcc{Information Fusion through EKF}
		\tcp{{Prediction}:}
		$\bm{\mu}^{\text{(A)}}_-[:,n]=\mathbf{F}^{\text{(A)}}\bm{\mu}^{\text{(A)}}_+[:,n-N_s^{\text{(A)}}]$\;\label{alg_EKF_RT_start}
		$\bm{\Sigma}^{\text{(A)}}_-[:,:,n]=\mathbf{F}^{\text{(A)}}\bm{\Sigma}^{\text{(A)}}_+[:,:,n-N_s^{\text{(A)}}]\left(\mathbf{F}^{\text{(A)}}\right)^{\text T} + \mathbf{C}^{\text{(A)}}_u$\;
		\tcp{{Update}:}
		${\bf{J}}^{\text{(A)}} = \frac{\partial \mathbf{h}^{\text{(A)}}({\bf{S}}[:,n])}{\partial {\bf{S}}^{\text T}[:,n]} \big|_{{\bf{S}}[:,n] = \bm{\mu}^{\text{(A)}}_-[:,n]}$\;
		$\mathbf{K}^{\text{(A)}}=\mathbf{\Sigma}^{\text{(A)}}_-[:,:,n]\left(\mathbf{J}^{\text{(A)}}\right)^{\text T}\big(\mathbf{J}^{\text{(A)}}\mathbf{\Sigma}^{\text{(A)}}_-[:,:,n]\left(\mathbf{J}^{\text{(A)}}\right)^{\text T}+\mathbf{C}^{\text{(A)}}_w\big)^{-1}$\;
		$\bm{\mu}^{\text{(A)}}_+[:,n]=\bm{\mu}^{\text{(A)}}_-[:,n]+\mathbf{K}^{\text{(A)}}\big(\mathbf{\hat X}^{\text{(A)}}[:,n]-\mathbf{h}^{\text{(A)}}(\bm{\mu}^{\text{(A)}}_-[:,n])\big)$\;
		$\mathbf{\Sigma}^{\text{(A)}}_+[:,:,n]=\big(\mathbf{I-K^{\text{(A)}}J^{\text{(A)}}}\big)\mathbf{\Sigma}^{\text{(A)}}_-[:,:,n]$.\label{alg_EKF_RT_end}
	\end{scriptsize}
\end{algorithm}

\vspace{-5mm}
\subsection{Proposed Dual-Compensation-Based Demodulation and Tracking for Passive Vehicle}\label{sec_EKF4CT}
\newcounter{TempEqCnt}
\setcounter{TempEqCnt}{\value{equation}}
\setcounter{equation}{48}
\begin{figure*}[b]
	\vspace{-2mm}
	\hrulefill
	\begin{footnotesize}
		\begin{equation}\label{eq_betaD}
			\bm{\hat{\beta}}^{\text{(D)}}[n]\! =\! W\! \left(\! \text{Avg}\left({\hat{\bf{A}}}\odot {\mathcal{\underline{Y}}}^{({\text{P}})}\left[ \left[\kern-0.30em\left[ \mathbf{r}^{\text{(P)}}[n]/r_{\text{res}}^{({\text{P}})} 
			\right]\kern-0.30em\right]\hspace*{-4mm} \mod {N_f},\left[\kern-0.30em\left[ \mathbf{v}^{\text{(P)}}[n]/v_{\text{res}}^{({\text{P}})}\! +\! 0:{{{N_c}}}/{{{N_{\text{tx}}}}}:{N_c}\! -\! {{{N_c}}}/{{{N_{\text{tx}}}}} 
			\right]\kern-0.30em\right]\hspace*{-4mm}\mod {N_c},: \right] \right)/{\bm{\hat{\beta}}[n-1]}\! \right)\! ,\!
		\end{equation}
	\end{footnotesize}
	\vspace{-8mm}
\end{figure*}
\setcounter{TempEqCnt}{\value{equation}}
\setcounter{equation}{49}
\begin{figure*}[b]
	\hrulefill
	\begin{footnotesize}
		\begin{equation}\label{eq_beta_n}\boldsymbol{\hat \beta}[n] = \text{Avg}\left({\hat{\bf{A}}}\odot {\mathcal{\underline{Y}}}^{({\text{P}})}\left[ \left[\kern-0.30em\left[ \mathbf{r}^{\text{(P)}}[n]/r_{\text{res}}^{({\text{P}})} 
			\right]\kern-0.30em\right]\hspace*{-4mm}\mod {N_f},\left[\kern-0.30em\left[ \mathbf{v}^{\text{(P)}}[n]/v_{\text{res}}^{({\text{P}})} + 0:{N_c}/{N_{\text{tx}}}:{N_c} - {N_c}/{N_{\text{tx}}} 
			\right]\kern-0.30em\right]\hspace*{-4mm}\mod {N_c},: \right] \right)/{\bm{\hat{\beta}}^{\text{(D)}}[n]} .
		\end{equation}
	\end{footnotesize}
	\vspace{-8mm}
\end{figure*}	
  The goal of JCR in the PV side is to track the AV and estimate other paths while demodulating the data modulated on the chirp waveform. The overall procedure is outlined as follows.
  First, the PV predicts the current AV's parameters based on the previous AV's parameters.
  Second, the current estimation results are obtained from Algorithm~\ref{alg_ParaEst}, which are used to represent the true parameters of AV and the payload data. The payload data is then extracted by comparing the current observation results with the predicted parameters. 
  Third, the current estimation results, excluding payload data, are used to represent the true parameters of AV.
  Finally, the fusion of the AV's parameters is obtained by combining the predictions based on the previous AV's parameters and the current estimation results excluding the payload data.
  The term ``dual-compensation'' stems from two perspectives: from the demodulation viewpoint, AV's parameters are compensated, while from the tracking perspective, modulated data is eliminated.

  The problem for the PV to solve can also be formulated as the 4D parameter estimation in (\ref{eq_problemDIR}) and the Bayes minimum mean square error in (\ref{eq_BMMSE}).
  Since the parameters estimated from the non-line-of-sight paths between the AV and PV do not represent the parameters of the scatterers to the PV, the PV only tracks and fuses the parameters of the AV. For other paths, PV only performs parameter estimation as described in Algorithm~\ref{alg_ParaEst}.
  The PV's observation vector, state vector, state equation, and observation equation are defined similar to those of the AV, as specified in (\ref{eq_ObsVec}) and (\ref{eq_StateEqObservationEq}), with ``(A)'' replaced by ``(P)''. Note that $N_s^{\text{(A)}}$ is replaced by $N_s^{\text{(P)}}\! =\! 1$, as the PV needs to demodulate data at each frame.
  The geometric relationship illustrating the relative motion between the AV and the PV, from the PV's perspective, is similar to that of Fig.~\ref{fig_geometry}. The detailed procedure for the PV to track the AV and to demodulate data is described as follows.

  The current distance and radial-velocity predicted by $\mathbf{\hat{S}}^{\text{(P)}}[:,n-1]\! =\! \big[\mathbf{\hat{c}}^{\text{(P)}}_x[n-1], \mathbf{\hat{c}}^{\text{(P)}}_y[n-1], \mathbf{\hat{v}}^{\text{(P)}}_x[n-1], \mathbf{\hat{v}}^{\text{(P)}}_y[n-1]\big]^{\text T}$, i.e., $\mathbf{\hat{r}}_p[n]$ and $\mathbf{\hat{v}}_{p,r}[n]$, are given by
\setcounter{TempEqCnt}{\value{equation}}
\setcounter{equation}{44}  
{\footnotesize\begin{align} 
	\mathbf{\hat{r}}_p[n] &= \bigg( \big({\mathbf{\hat{c}}^{\text{(P)}}_x}[n - 1] + {\mathbf{\hat{v}}^{\text{(P)}}_x}[n - 1]{{\bar T}_c}{N_c}\big)^2 + \big({\mathbf{\hat{c}}^{\text{(P)}}_y}[n - 1] \nonumber\\
	&+ {\mathbf{\hat{v}}^{\text{(P)}}_y}[n - 1]{{\bar{T}}_c}{N_c}\big)^2 \bigg)^{0.5}, \label{eq_r_n} \\
	\mathbf{\hat{v}}_{p,r}[n] &= {\mathbf{\hat{v}}^{\text{(P)}}_x}[n - 1] \sin \big(\hat{\theta}_p\big) + {\mathbf{\hat{v}}^{\text{(P)}}_y}[n - 1] \cos \big(\hat{\theta}_p \big), \label{eq_v_n}   
\end{align}}
where $\hat{\theta}_{p}\! =\! {\arcsin}\Big({\mathbf{\hat{c}}^{\text{(P)}}_x}[n - 1]/ \big(\big(\mathbf{\hat{c}}_x^{\text{(P)}}[n - 1]\big)^2 + \big(\mathbf{\hat{c}}_y^{\text{(P)}}[n - 1]\big)^2\big)^{0.5} \Big)$.
  The current observation vector $\mathbf{\hat{X}}^{\text{(P)}}[:,n]\! =\! \big[\mathbf{\hat{r}}^{\text{(P)}}[n],\mathbf{\hat{v}}^{\text{(P)}}_r[n], \bm{\hat{\theta}^{\text{(P)}}}[n]\big]^{\text T}$ modulated with data is obtained using Algorithm~\ref{alg_ParaEst}.

  With the distance and radial-velocity predicted by the last frame and the parameters estimated from the current frame, $\mathbf{\hat{f}}_r^{\text{(D)}}[n]$ and $\mathbf{\hat{f}}_v^{\text{(D)}}[n]$ are calculated by
{\footnotesize\begin{align} 
	\mathbf{\hat{f}}_r^{{\text{(D)}}}[n] =& \left[\kern-0.30em\left[ \big(\mathbf{\hat{r}}^{\text{(P)}}[n] - \mathbf{\hat{r}}_p[n]\big)/r_{\text{res}}^{({\text{P}})} \right]\kern-0.30em\right]\hspace*{-2mm} \mod ({N_f}/2), \label{eq_f_r_D} \\
	\mathbf{\hat{f}}_v^{{\text{(D)}}}[n] =& \left\{ \begin{array}{*{20}{c}}
		\left[\kern-0.30em\left[ \frac{{\mathbf{\hat{v}}^{\text{(P)}}_r}[n] - {\mathbf{\hat{v}}}_{p,r}[n]}{v_{\text{res}}^{({\text{P}})}} \right]\kern-0.30em\right]\hspace*{-2mm} \mod ({N_c}),\;\text{if}\;{\text{beacon}}\;{\text{frame}}, \\
	\left[\kern-0.30em\left[ \frac{{\mathbf{\hat{v}}^{\text{(P)}}_r}[n] - {\mathbf{\hat{v}}}_{p,r}[n]}{v_{\text{res}}^{({\text{P}})}} \right]\kern-0.30em\right]\hspace*{-2mm} \mod \big(\frac{N_c}{N_{\text{tx}}}\big),\;\text{if}\;{\text{DDM}}\;{\text{frame}} .
	\end{array} \right.  \label{eq_f_v_D}
\end{align}}
After $\mathbf{\hat{f}}_r^{\text{(D)}}[n]$ and $\mathbf{\hat{f}}_v^{\text{(D)}}[n]$ are obtained, $\bm{\hat{\beta}}^{\text{(D)}}[n]$ is computed according to (\ref{eq_betaD}) at the bottom of this page,
where $W(\cdot)$ is the function to find which QAM symbol has the minimum distance to the input, $\bm{\hat{\beta}}[n-1]$ is the estimated complex amplitude of the last frame, and $\hat{\bf{A}}\! \in\! \mathbb{C}^{N_{\text{tx}}\times N_{\text{rx}}}$ is the conjugate estimated array response matrix given by
\setcounter{TempEqCnt}{\value{equation}}
\setcounter{equation}{50}
\begin{small}
\begin{align}\label{eqCEAR} 
	&\hat{\bf{A}}[n_{\text{tx}},n_{\text{rx}}]\nonumber\\
	&= e^{-\textsf{j}2\pi \frac{{{f_c}}}{c}\big(({n_{\text{tx}}}\!\!\!\! \mod {N_{\text{tx},a}}){d_{\text{tx},a}}\cos \bm{\hat{\varphi}}_{\text{tx}}^{(\text{P})} [n]\sin \bm{\hat{\theta}}_{\text{tx}}^{(\text{P})} [n] + \left\lfloor {\frac{{{n_{\text{tx}}}}}{{{N_{\text{tx},a}}}}} \right\rfloor {d_{\text{tx},e}}\sin \bm{\hat{\varphi}}_{\text{tx}}^{(\text{P})} [n]\big)} \nonumber \\
	& \times e^{-\textsf{j}2\pi \frac{{{f_c}}}{c}\big(({n_{\text{rx}}}\!\!\!\! \mod {N_{\text{rx},a}}){d_{\text{rx},a}}\cos \bm{\hat{\varphi}}_{\text{rx}}^{(\text{P})} [n]\sin \bm{\hat{\theta}}_{\text{rx}}^{(\text{P})} [n] + \left\lfloor {\frac{{{n_{\text{rx}}}}}{{{N_{\text{rx},a}}}}} \right\rfloor {d_{\text{rx},e}}\sin \bm{\hat{\varphi}}_{\text{rx}}^{(\text{P})} [n]\big)} ,
\end{align}
\end{small}
in which $\bm{\hat{\theta}}_{\text{tx}}^{(\text{P})}[n]$, $\bm{\hat{\varphi}}_{\text{tx}}^{(\text{P})}[n]$, $\bm{\hat{\theta}}_{\text{rx}}^{(\text{P})}[n]$, and $\bm{\hat{\varphi}}_{\text{rx}}^{(\text{P})} [n]$ are the estimated AoD of the AV, EoD of the AV, AoA of the PV, and EoA of the PV, respectively, and they are obtained by Algorithm~\ref{alg_ParaEst}.
  Then $\bm{\hat{\beta}}[n]$ is updated according to (\ref{eq_beta_n}) at the bottom of this page.
Finally, the observation vector $\mathbf{\hat{X}}^{\text{(P)}}[:,n]\! =\! \left[\mathbf{\hat{r}}^{\text{(P)}}[n], \mathbf{\hat{v}}^{\text{(P)}}_r[n], \bm{\hat{\theta}^{\text{(P)}}}[n]\right]^{\text T}$ with the modulated data removed is given by 
\begin{equation}\label{eq_RmDataCT} 
	\mathbf{\hat{X}}^{\text{(P)}}[:,n] \Leftarrow \mathbf{\hat{X}}^{\text{(P)}}[:,n] - \left[\mathbf{\hat{f}}_r^{{\text{(D)}}}[n]r_{\text{res}}^{\text{(P)}}, \mathbf{\hat{f}}_v^{{\text{(D)}}}[n]v_{\text{res}}^{\text{(P)}}, 0\right]^{\text T}.
\end{equation}
After the modulated data is removed, the EKF algorithm is utilized to perform tracking \cite{ref_StatisticalSP}, as shown in steps~\ref{alg_DDF_EKF_start}--\ref{alg_DDF_EKF_end} of Algorithm~\ref{alg_EKF_CT}.
  The definitions of the variables are similar to those given in Section~\ref{sec_EKF4RT}. The posterior mean and variance first time used in Algorithm~\ref{alg_EKF_CT} are derived from the initial channel sensing stage.
 
\begin{algorithm}[!t]
	\label{alg_EKF_CT} 
	\begin{scriptsize}
		\caption{Proposed Dual-Compensation-Based Demodulation and Tracking for PV}
		\LinesNumbered 
		\KwIn{Posterior mean vector $\bm{\mu}^{\text{(P)}}_+[:,n-1]\! \in\! \mathbb{R}^{4}$ and posterior covariance matrix $\bm{\Sigma}^{\text{(P)}}_+[:,:,n-1]\!\in\! \mathbb{R}^{4\times 4}$ of state vector $\mathbf{\hat{S}}^{\text{(P)}}[:,n-1]\! \in\! \mathbb{R}^{4}$, state perturbation covariance matrix $\mathbf{C}^{\text{(P)}}_u\! \in\! \mathbb{R}^{4\times 4}$ and observation noise covariance matrix $\mathbf{C}^{\text{(P)}}_w\! \in\! \mathbb{R}^{4\times 4}$, received signal after mixing $\tilde{\mathcal{Y}}^{({\text{P}})}\! \in\! \mathbb{C}^{N_f\times N_c\times N_{\text{rx}}}$ (or ${\mathcal{Y}}^{({\text{P}})}\! \in\! \mathbb{C}^{N_f\times N_c\times N_{\text{rx}}}$); }
		\KwOut{Demodulated data $\mathbf{\hat{f}}_r^{\text{(D)}}[n]$, $\mathbf{\hat{f}}_v^{\text{(D)}}[n]$ and $\bm{\hat{\beta}}^{\text{(D)}}[n]$, posterior mean vector $\bm{\mu}^{\text{(P)}}_+[:,n]\! \in\! \mathbb{R}^{4}$ and posterior covariance matrix $\bm{\Sigma}^{\text{(P)}}_+[:,:,n]\! \in\! \mathbb{R}^{4\times 4}$ of current state vector $\mathbf{S}^{\text{(P)}}[:,n]\! \in\! \mathbb{R}^{4}$;}
		Perform DFT on ${\tilde{\mathcal{Y}}^{({\text{P}})}}$ (or ${{\mathcal{Y}}^{({\text{P}})}}$) to obtain ${\tilde{\mathcal{\underline Y}}^{({\text{P}})}}$ (or ${{\mathcal{\underline Y}}^{({\text{P}})}}$) as (\ref{eq_2D_DFT})\;
		Obtain $\tilde{\bf{Y}}_{{\text{RDM}}}^{({\text{P}})}$ (or ${\bf{Y}}_{{\text{RDM}}}^{({\text{P}})}$)  as (\ref{eq_RDM})\;
		Perform CFAR detection on $\tilde{\bf{Y}}_{{\text{RDM}}}^{({\text{P}})}$ (or ${\bf{Y}}_{{\text{RDM}}}^{({\text{P}})}$) to obtain ${\tilde{\bf{Y}}^{({\text{P}})}_{\text{RDM,D}}}$ (or ${{\bf{Y}}^{({\text{P}})}_{\text{RDM,D}}}$)\;
		Predict $\mathbf{\hat{r}}^{\text{(P)}}[n]$, $\mathbf{\hat{v}}^{\text{(P)}}_r[n]$ as $\mathbf{\hat{r}}_p[n]$, $\mathbf{\hat{v}}_{p,r}[n]$ according to $\bm{\mu}^{\text{(P)}}_+[:,n-1]$ based on (\ref{eq_r_n}), (\ref{eq_v_n})\;\label{DDF_predict}
		Obtain observation vector $\mathbf{X}^{\text{(P)}}[:,n]\! =\! \big[\mathbf{\hat{r}}^{\text{(P)}}[n], \mathbf{\hat{v}}^{\text{(P)}}_r[n], \bm{\hat{\theta}}^{\text{(P)}}[n]\big]^{\text T}$ based on ${\tilde{\bf{Y}}^{({\text{P}})}_{\text{RDM,D}}}$ (or ${{\bf{Y}}^{({\text{P}})}_{\text{RDM,D}}}$) through Algorithm~\ref{alg_ParaEst}\;\label{alg_3_observation}		
		Demodulate data modulated in distance and velocity, i.e., $\mathbf{\hat{f}}_r^{\text{(D)}}[n]$ and $\mathbf{\hat{f}}_v^{\text{(D)}}[n]$, by comparing $\mathbf{X}^{\text{(P)}}[:,n]$ with $\mathbf{\hat{r}}_p[n]$ and $\mathbf{\hat{v}}_{p,r}[n]$ as (\ref{eq_f_r_D}) and (\ref{eq_f_v_D})\;
		Demodulate data modulated in amplitude, i.e., $\bm{\hat{\beta}}^{\text{(D)}}[n]$, as (\ref{eq_betaD})\;
		Remove all data modulated in $\mathbf{X}^{\text{(P)}}[:,n]$ as (\ref{eq_RmDataCT})\;
		\tcc{Information Fusion of \textbf{the AV} through EKF}
		\tcp{{Prediction}:}
		$\bm{\mu}^{\text{(P)}}_-[:,n]=\mathbf{F}^{\text{(P)}}\bm{\mu}^{\text{(P)}}_+[:,n-1]$\;\label{alg_DDF_EKF_start}
		$\bm{\Sigma}^{\text{(P)}}_-[:,:,n]=\mathbf{F}^{\text{(P)}}\bm{\Sigma}^{\text{(P)}}_+[:,:,n-1]\left(\mathbf{F}^{\text{(P)}}\right)^{\text T} + \mathbf{C}^{\text{(P)}}_u$\;
		\tcp{{Update}:}
		${\bf{J}}^{\text{(P)}} = \frac{\partial \mathbf{h}^{\text{(P)}}({\bf{S}}[:,n])}{\partial {\bf{S}}^{\text T}[:,n]} \big|_{{\bf{S}}[:,n] = \bm{\mu}^{\text{(P)}}_-[:,n]}$\;
		$\mathbf{K}^{\text{(P)}}=\mathbf{\Sigma}^{\text{(P)}}_-[:,:,n]\left(\mathbf{J}^{\text{(P)}}\right)^{\text T}\left(\mathbf{J}^{\text{(P)}}\mathbf{\Sigma}^{\text{(P)}}_-[:,:,n]\left(\mathbf{J}^{\text{(P)}}\right)^{\text T}+\mathbf{C}^{\text{(P)}}_w\right)^{-1}$\;
		$\bm{\mu}^{\text{(P)}}_+[:,n]=\bm{\mu}^{\text{(P)}}_-[:,n]+\mathbf{K}^{\text{(P)}}(\mathbf{X}^{\text{(P)}}[:,n]-\mathbf{h}^{\text{(P)}}(\bm{\mu}^{\text{(P)}}_-[:,n]))$\;
		$\mathbf{\Sigma}^{\text{(P)}}_+[:,:,n]=(\mathbf{I-K^{\text{(P)}}J^{\text{(P)}}})\mathbf{\Sigma}^{\text{(P)}}_-[:,:,n]$. \label{alg_DDF_EKF_end}
	\end{scriptsize}
\end{algorithm}

\subsection{Complexity Analysis}
The computational complexities of Algorithm~\ref{alg_EKF_RT} and Algorithm~\ref{alg_EKF_CT} are both dominated by the parameter estimation routine detailed in Algorithm~\ref{alg_ParaEst}. The additional steps in each algorithm introduce a lower-order computational overhead, which we analyze as follows.

For Algorithm~\ref{alg_EKF_RT}, the additional overhead stems from two main operations. First, the data pre-compensation step (\ref{eq_RmDataRT}) involves element-wise operations on the received signal tensor, leading to a complexity of $\mathcal{O}(N_{\text{rx}}N_fN_c)$. Second, the EKF tracking stage operates on fixed, low-dimensional state (4D) and observation (3D) vectors, resulting in a negligible constant complexity of $\mathcal{O}(1)$ per target.

For Algorithm~\ref{alg_EKF_CT}, the additional steps consist of data demodulation and tracking. The prediction-based demodulation for delay and Doppler ((\ref{eq_r_n})--(\ref{eq_f_v_D})) has a constant complexity of $\mathcal{O}(1)$. The demodulation of the QAM amplitude symbol ((\ref{eq_betaD})--(\ref{eqCEAR})) incurs a complexity of $\mathcal{O}(N_{\text{rx}} N_c / N_{\text{tx}})$. The final EKF update, similar to the AV's case, contributes a constant complexity of $\mathcal{O}(1)$.

In both algorithms, the total complexity is dictated by the initial parameter estimation, while the subsequent tracking and demodulation steps do not significantly increase the overall computational load.

\vspace{-2mm}
\section{Simulation Results}\label{Sec_Simulation}
\begin{table}[!t]
	\vspace{-5mm}
	\scriptsize
	\centering
	\caption{Simulation Parameters}
	\label{table_para}
	\vspace*{-3mm}
	\begin{tabular}{lll|lll}
		\hline\hline
		Parameter & Value & Unit   &   Parameter & Value & Unit \\ \hline\hline
		$f_c$             & 80 & GHz   &    $T_c$                           & $51.2$              & $\mu s$ \\
		$B$               & 640     & MHz   &    $\bar{T}_c$                     & $51.2\times68/60$ & $\mu s$ \\
		$f_s$             & 20    & MHz   &    $\tilde{T}_c$                   & $51.2\times64/60$ & $\mu s$ \\
		$N_c$             & 128   & NA    &    $T_g$                           & $50\times4/60$   & $\mu s$ \\
		$N_f$             & 1024   & NA    &    $T_{\text{sen}}=T_{\text{com}}$ & $51.2/60$          & $\mu s$ \\
		$N_{\text{tx}}$   & 4     & NA    &    $N_{\text{rx},a}$               & 8                 & NA \\
		$N_{\text{rx}}$   & 16    & NA    &    $N_{\text{rx},e}$               & 2                 & NA \\
		$N_{\text{tx},a}$ & 2     & NA    &    $\theta_{\text{max}}$           & 60                & degree \\
		$N_{\text{tx},e}$ & 2     & NA    &    $\varphi_{\text{max}}$          & 15                & degree \\ \hline \hline
	\end{tabular}
	\vspace{-5mm}	
\end{table}
\subsection{System Parameters and Performance Metrics}\label{Sec_SysPara}

\subsubsection{System Setup}

  The key system parameters are listed in Table~\ref{table_para}.
  The time parameters in Table~\ref{table_para} can be obtained based on the relationships shown in Fig.~\ref{fig_chirp}. The detailed calculation process is omitted here due to space limitations.
  The other system parameters are given as follows. $\varpi_r\! =\! \varpi_v\! =\! 1$, $N_s^{\text{(A)}}\! =\! 10$, $N_s^{\text{(P)}}\! =\! 1$, $N_Q\! =\! 4,16,64$, i.e., QPSK, 16QAM, and 64QAM, false alarm rate is $0.001$, $N_o\! =\! 10$, $\lambda\! =\! c/f_c\! =\! 0.0037$\,m, $d_{\text{rx},a}\! =\! 0.5774\lambda$, and $d_{\text{rx},e}\! =\! 1.9319\lambda$.
  The noise power spectrum density at the receiver is set to $\sigma_{\text{NSD}}^2\! =\! -174$\,dBm/Hz.
  The complex channel coefficient of the two-way path is calculated as $\beta_h\! =\! \beta_a e^{\textsf{j}\xi}\sqrt{G_{\text{tx}}G_{\text{rx}}S_{\mathrm{eff}}\lambda^{2}}/\sqrt{\left(4\pi\right)^{3}(R_{1})^{2}(R_{2})^{2}}$,
where $\beta_a$ is the atmospheric attenuation and is set to $2$\,dB/km \cite{refITU}, $S_{\mathrm{eff}}$ is the effective reflection area of targets and is set to 4 in the simulations, $\xi\! \in\! \mathcal{U}(0,2\pi)$ is the random phase shift of the channel, while $G_{\text{tx}}$ and $G_{\text{rx}}$ are the antenna gains of the transmit antenna and receive antenna, respectively, and they are set to 1 in the simulations, $R_1$ and $R_2$ are the distance from the transmitter to the target and the distance from the target to the receiver, respectively.
  The complex channel coefficient of the one-way path can be calculated as $\beta_h\! =\! \beta_a e^{\textsf{j}\xi}\sqrt{G_{\text{tx}}G_{\text{rx}}\lambda^2}/4\pi R$, where $R$ is the distance from the transmitter to the receiver.
  In the tracking simulation, 
the transmit power is fixed at 5\,dBm. Hanning window is chosen as the window function for $\mathbf{w}_f$ and $\mathbf{w}_c$.

  The system environment is configured as follows. The coordinates of the AV (vehicle 1), the PV (vehicle 2) and the other vehicle (vehicle 3) are $(0, 0, 1)$\,m, $(-5, 5, 1)$\,m and $(5, -10, 1)$\,m, respectively.
  The velocities of the AV, the PV and the vehicle 3 are 20\,m/s, 25\,m/s, and 30\,m/s, respectively, and they are oriented along the positive $y$-axis.

\subsubsection{Performance Metrics}

  We adopt the hitrate as the metric to evaluate the accuracy of the parameter estimation \cite{ref_JSTSP_DingyouMa_IM}. A ``hit'' refers to the target (path) being detected. The aim of our scheme is to estimate the parameters of the target (path) within an acceptable margin of error, rather than pursuing the theoretically optimal outcome. A hit happens when 
\begin{align}\label{eqHit} 
	& |\mathbf{\hat{r}}^{(\cdot)}[n] - \mathbf{r}^{(\cdot)}[n]|\le r_{\text{res}}^{(\cdot)}\ \&\ 
	|\mathbf{\hat{v}}_r^{(\cdot)}[n] - \mathbf{v}_r^{(\cdot)}[n]|\le v_{\text{res}}^{(\cdot)}.
\end{align}	
Let the total number of simulation runs be $N_{\text{tot}}$ and the number of hits therein be $N_{\text{hit}}$. Then the hitrate is $N_{\text{hit}}/N_{\text{tot}}$. We also use the cumulative distribution function (CDF) to evaluate the sensing accuracy of the proposed beacon frame-aided 4D-parameter estimation scheme.
  As information fusion is a continuous-time process, the evaluation of tracking performance is essential. We present the true values, estimated values, and fused values simultaneously to illustrate the tracking process. Additionally, the tracking performance is also evaluated by analyzing the errors between the true and estimated values of orientation, velocity in the $x$-direction, and velocity in the $y$-direction. Notably, these three parameters cannot be obtained through conventional schemes without the addition of extra hardware\cite{ref_TAES_HorizontalVel}.
  Furthermore, the symbol error rate (SER) performance is presented to demonstrate the communication capabilities of the DD-QAM.

\vspace{-3mm}
\subsection{Detection, Estimation and Demodulation Performance Evaluation}\label{S7.2}

\begin{figure*}[!t]
	\subfigure[]{
		\begin{minipage}[t]{0.33\linewidth}
			\includegraphics[width=2.2in]{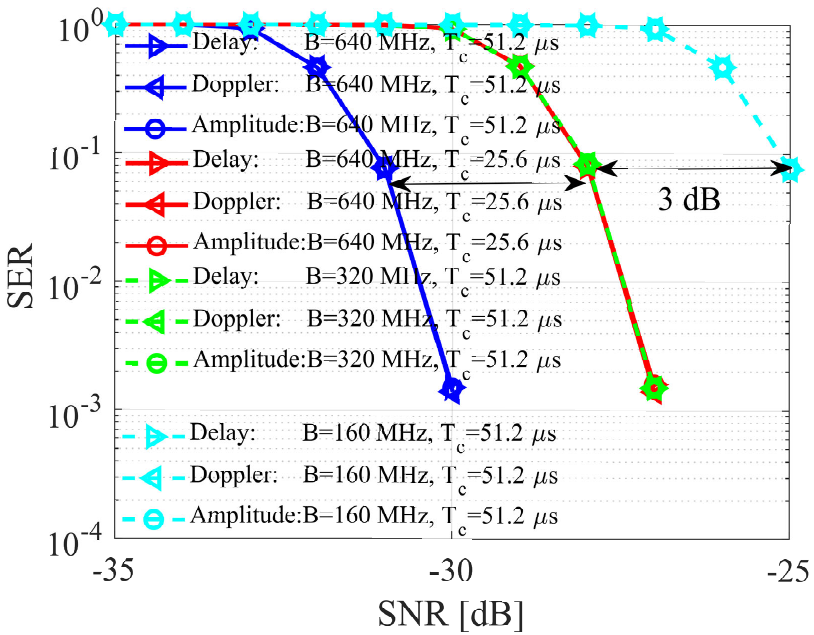}
			\label{fig_BER} 
		\end{minipage}%
	}%
	\subfigure[]{
		\begin{minipage}[t]{0.33\linewidth}
			\includegraphics[width=2.2in]{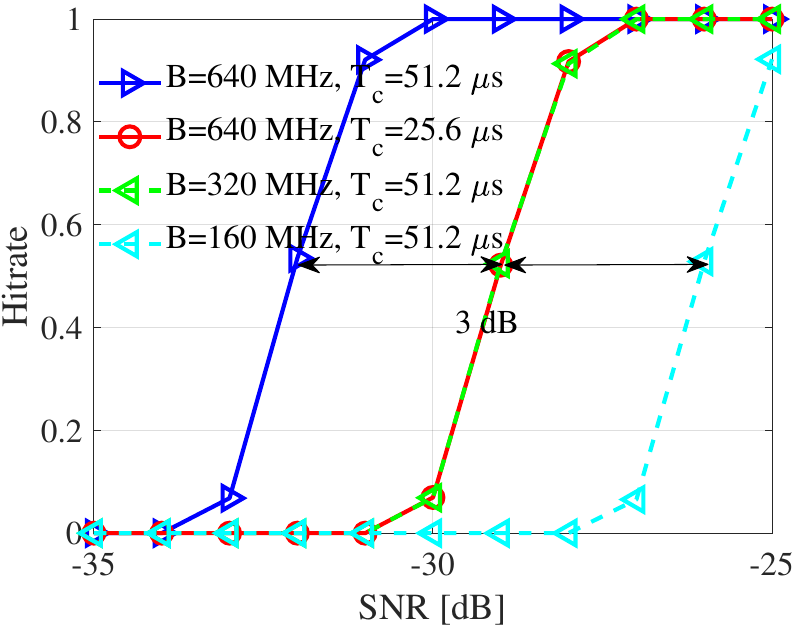}
			\label{fig_Hitrate} 
		\end{minipage}%
	}
	\subfigure[]{
		\begin{minipage}[t]{0.33\linewidth}
			\includegraphics[width=2.2in]{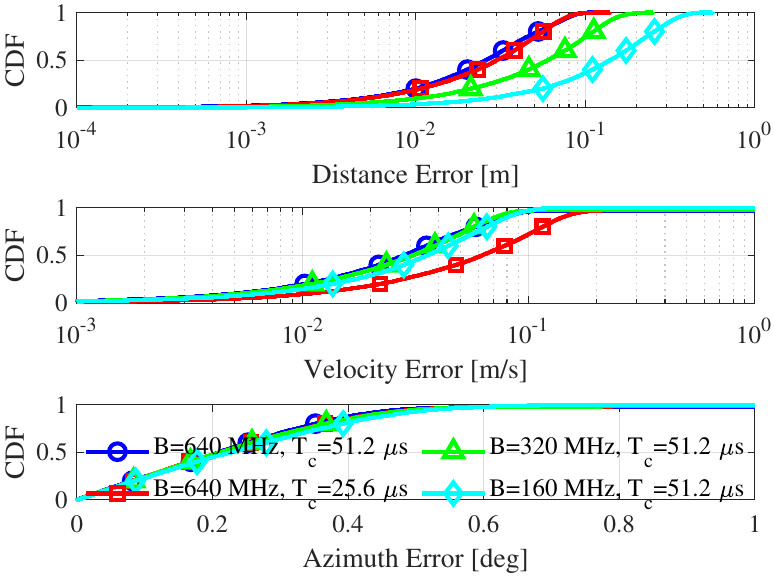}
			\label{fig_CDF_SNR} 
		\end{minipage}%
	}%
	\centering
	\vspace{-3mm}	
	\caption{Detection, estimation and demodulation performance under different values of bandwidth $B$, effective sampling time $T_c$, and SNR settings:
		(a) SER;
		(b) hitrate;
		(c) CDF.
		(a)~is the PV's data demodulation performance,
		(b) and (c) are the AV's parameter estimation performance in estimating the distance, radial-velocity, and AoA of the PV relative to itself.
		The SNR in (c) is set to $-25$\,dB.
	}
	\label{fig_SenComm} 
	\vspace{-3mm}
\end{figure*}

  We fix the state of the environment as the initial state as specified in Section~\ref{Sec_SysPara}, and iterate 10000 times to obtain results shown in Fig.~\ref{fig_SenComm}.
  Fig.~\ref{fig_BER} shows the demodulation performance of the PV, and several observations are noteworthy.
First, the reason for the worst SER being 1 is that if the detection fails, no demodulation takes place.
  Second, SERs derived from delay estimation, Doppler estimation, and complex amplitude estimation are identical.
This indicates that successful detection leads to successful demodulation, and there is no case where the detection is successful but the demodulation fails.
  Third, when $T_c$ is fixed at $50$ $\mu s$, and $B$ varies from $640$ MHz to $320$ MHz and then to $160$ MHz, the required signal-to-noise ratio (SNR) increases gradually by $3$ dB at the same SER. This is due to our fixed ratio of $f_s/B$ at $1/32$. With a larger bandwidth $B$, more sampling points are obtained, resulting in an enhanced equivalent SNR through coherent superposition.
\begin{figure*}[!t]
	\vspace{-3mm}
	\subfigure[]{
		\begin{minipage}[t]{0.33\linewidth}
			\includegraphics[width=2.2in]{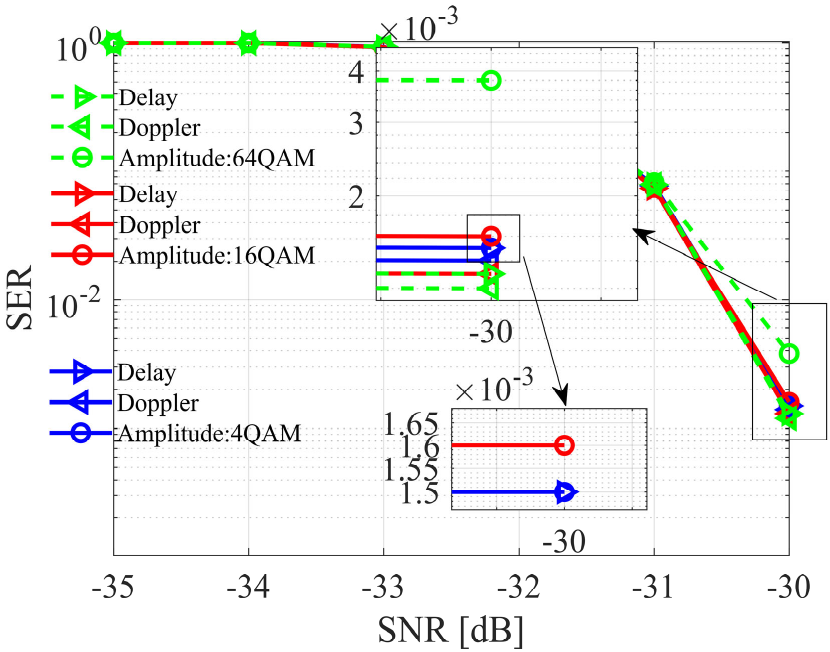}
			\label{fig_BER_QAM} 
		\end{minipage}%
	}%
	\subfigure[]{
		\begin{minipage}[t]{0.33\linewidth}
			\includegraphics[width=2.2in]{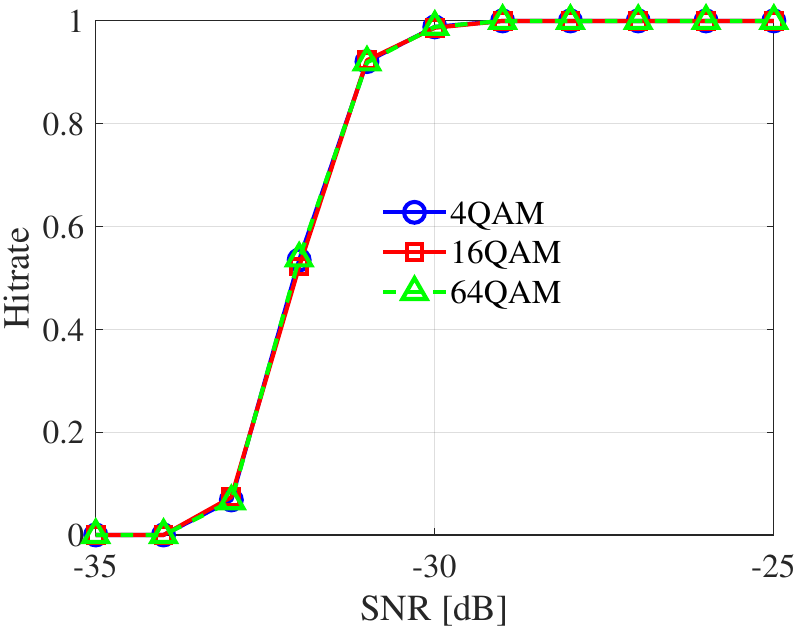}
			\label{fig_Hitrate_QAM} 
		\end{minipage}%
	}
	\subfigure[]{
		\begin{minipage}[t]{0.33\linewidth}
			\includegraphics[width=2.2in]{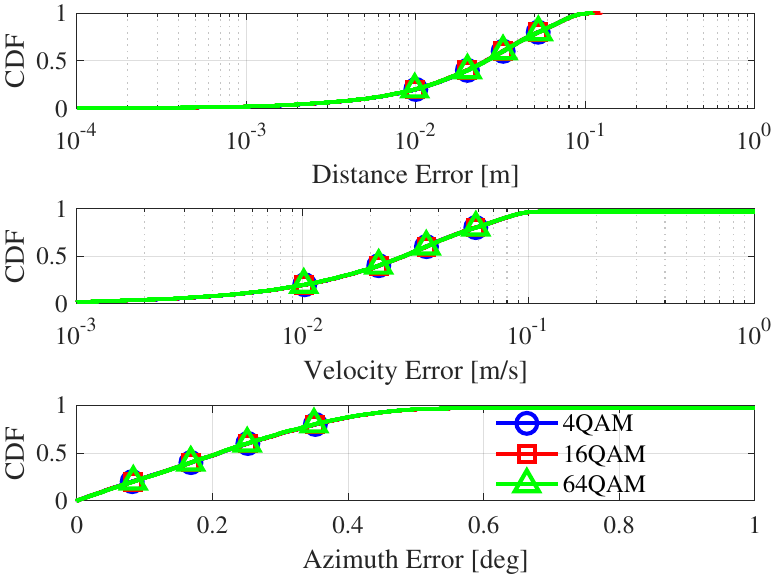}
			\label{fig_CDF_QAM} 
		\end{minipage}%
	}%
	\centering
	\vspace{-3mm}	
	\caption{Detection, estimation and demodulation performance with various QAM orders:
		(a) SER;
		(b) hitrate;
		(c) CDF.
		The SNR in (c) is set to $-25$\,dB.
	}
	\label{fig_SenComm_QAM} 
	\vspace{-3mm}
\end{figure*}

  Fig.~\ref{fig_Hitrate} illustrates the detection performance of the AV with hitrate as the metric. Since the success of detection means the success of data demodulation, the curves of Fig.~\ref{fig_Hitrate} correspond one-to-one with those of Fig.~\ref{fig_BER}. This correspondence resonates with the concept of the proposed DD-QAM, as data is modulated on ``parameters''.

  Fig.~\ref{fig_CDF_SNR} further examines the parameter estimation performance of Fig.~\ref{fig_Hitrate} with the SNR set at $-25$\,dB.
  Only successful detections in Fig.~\ref{fig_Hitrate} are considered in the statistics of Fig.~\ref{fig_CDF_SNR}.
We opt for a search precision of 1$^\circ$ in AoA to strike a balance between accuracy and computational complexity.

\begin{figure*}[!t]
  	\vspace{-6mm}
  	\centering
  	\subfigure[]{
  		\begin{minipage}[t]{0.5\linewidth}
  			\centering
  			\includegraphics[width=2.75in]{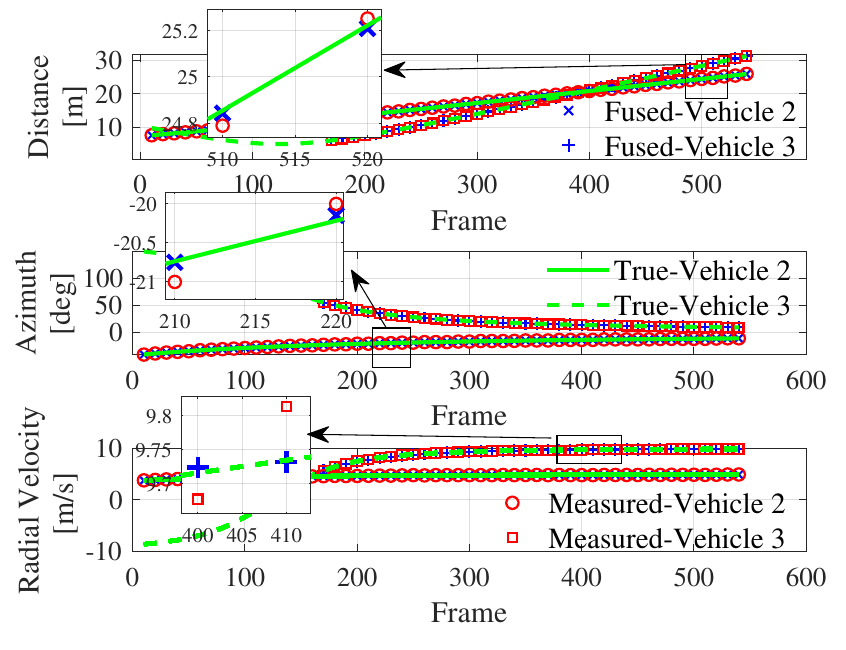}
  			\label{fig_RT_4D} 
  		\end{minipage}%
  	}%
  	\subfigure[]{
  		\begin{minipage}[t]{0.5\linewidth}
  			\centering
  			\includegraphics[width=2.5in]{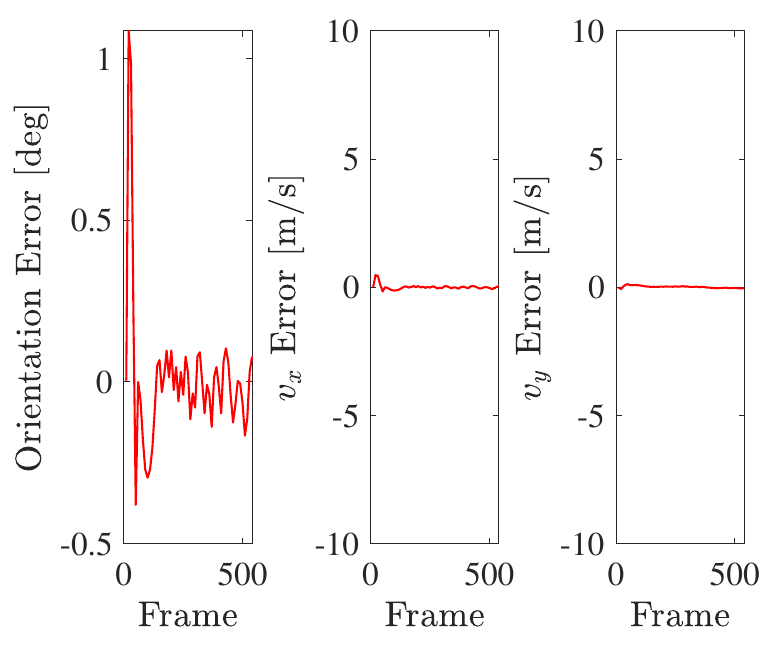}
  			\label{fig_RT_Orientation} 
  		\end{minipage}%
  	}%
  	\centering
  	\vspace{-2mm}	
  	\caption{AV(vehicle 1)'s tracking performance for PV (vehicle 2) and vehicle 3: (a)~distance, azimuth, and radial-velocity tracking performance;
  		(b)~errors in AV's estimation of PV's orientation and $x/y$-direction velocities.
  	}
  	\label{fig_RT} 
  	\vspace{-6mm}
  \end{figure*}

\begin{figure*}[!t]
  		\vspace{-2mm}
  	\subfigure[]{
  		\begin{minipage}[t]{0.5\linewidth}
  			\centering
  			\includegraphics[width=2.75in]{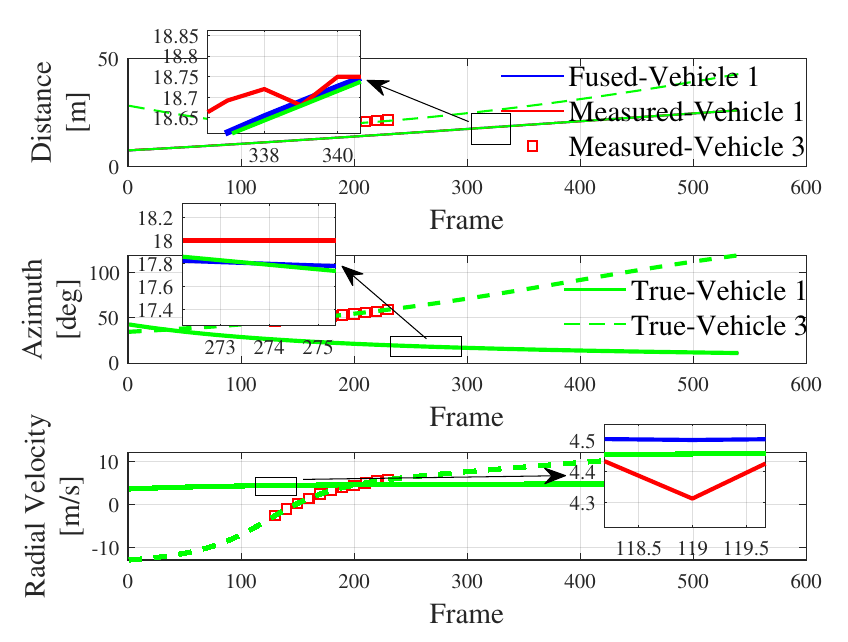}
  			\label{fig_CT_4D} 
  		\end{minipage}%
  	}%
  	\subfigure[]{
  		\begin{minipage}[t]{0.5\linewidth}
  			\centering
  			\includegraphics[width=2.5in]{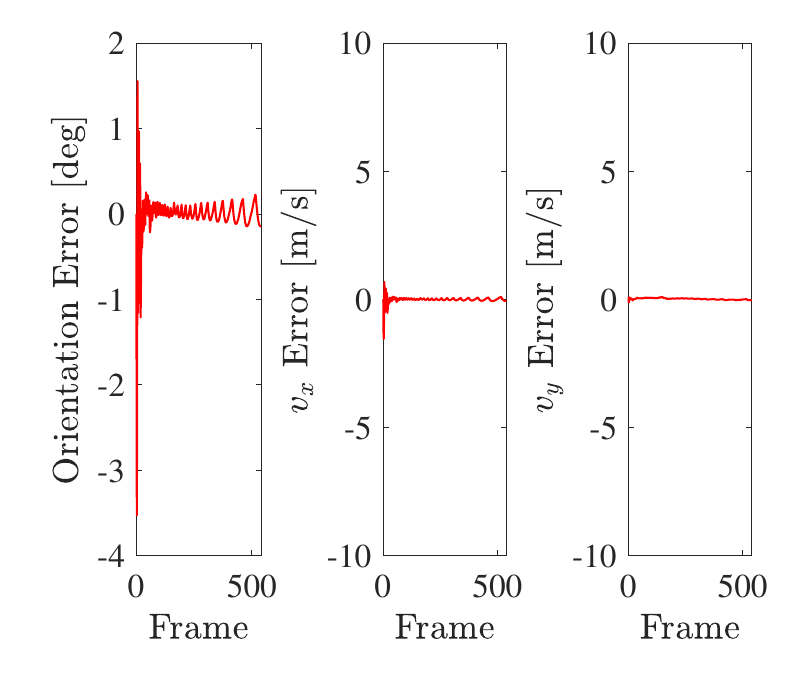}
  			\label{fig_CT_Orientation} 
  		\end{minipage}%
  	}%
  	\centering
  	\vspace{-2mm}	
  	\caption{PV (vehicle 2)'s tracking performance for AV (vehicle 1) and the path associated with vehicle 3: (a)~distance, azimuth, and radial-velocity tracking performance;
  		(b)~errors in PV's estimation of AV's orientation and $x/y$-direction velocities.
  	}
  	\label{fig_CT} 
  	\vspace{-4mm}
  \end{figure*}

  It can be seen that the performance of AoA estimation remains relatively consistent across different settings of bandwidth $B$ and effective sampling time $T_c$.
  By contrast, the accuracy of the delay and radial-velocity estimation improves significantly with either increased bandwidth or longer observation time. 
  This improvement arises from the improved delay resolution provided by larger bandwidth and the improved velocity resolution enabled by longer observation time.
  Notably, Algorithm~\ref{alg_ParaEst} uses arithmetic averaging of positions in the RDM affected by spectrum leakage. This approach results in estimation errors for distance and radial-velocity that are significantly lower than the resolutions presented in Table~\ref{table_SensingBound}.
  In fact, we can achieve off-grid parameter estimation during tracking with sequential estimation, thereby alleviating the computational burden on each frame. This will be verified in Figs.~\ref{fig_RT} and \ref{fig_CT}. 
  
  Fig.~\ref{fig_SenComm_QAM} illustrates the detection, estimation, and demodulation performance for different QAM orders.
  As shown in Fig.~\ref{fig_BER_QAM}, the QAM order only affects the demodulation performance—the higher the order, the worse the SER performance.
  The QAM order has no significant impact on detection and parameter estimation performance.
 \begin{figure*}[!t]
		\vspace{-3mm}
	\centering
	\subfigure[]{
		\begin{minipage}[t]{0.33\linewidth}
			\centering
			\includegraphics[width=2.2in]{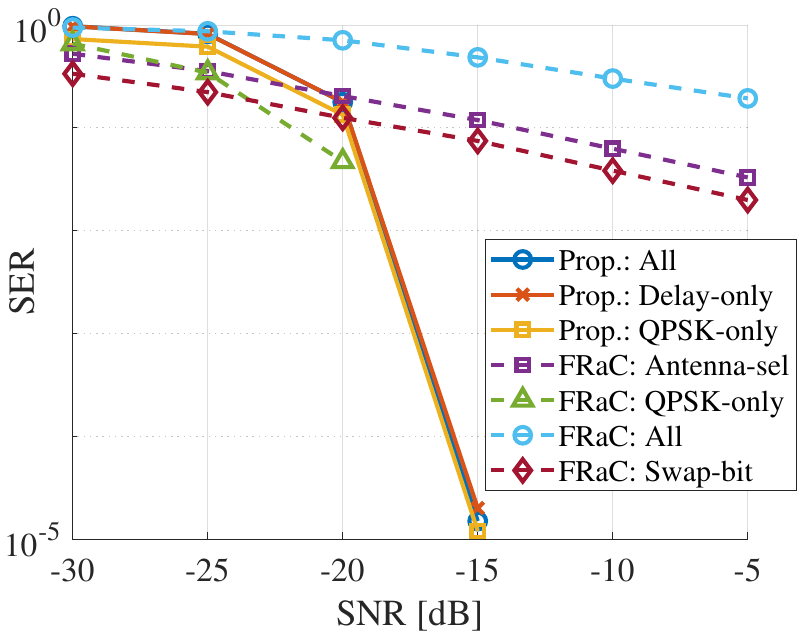}
			\label{fig_Cmp} 
		\end{minipage}%
	}%
	\subfigure[]{
		\begin{minipage}[t]{0.33\linewidth}
			\centering
			\includegraphics[width=2.2in]{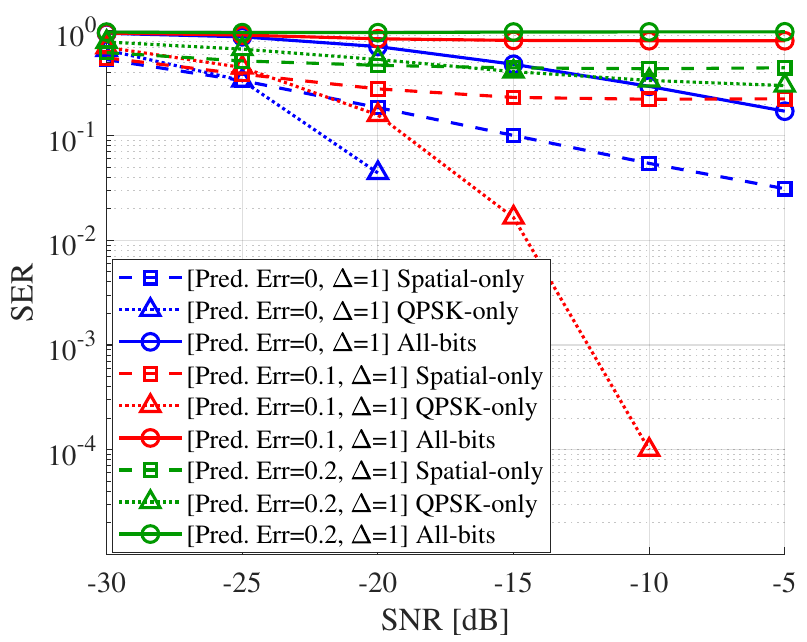}
			\label{fig_FRaC} 
		\end{minipage}%
	}%
	\subfigure[]{
		\begin{minipage}[t]{0.33\linewidth}
			\centering
			\includegraphics[width=2.2in]{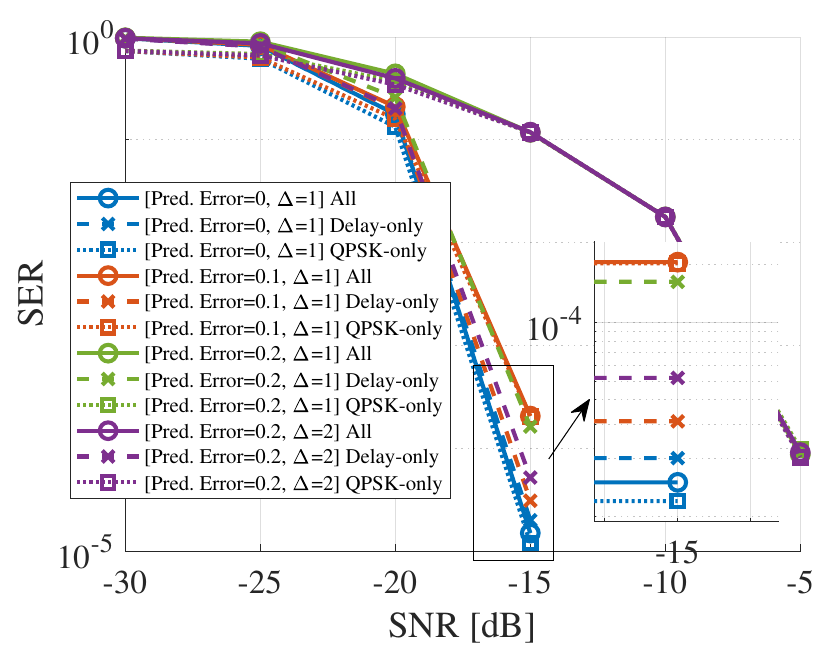}
			\label{fig_Mine} 
		\end{minipage}%
	}%
	\centering  	
	\vspace{-2mm}	
	\caption{SER performance of FRaC and proposed TDM-DD-QAM: (a) FRaC versus TDM-DD-QAM;
			(b) FRaC with prediction errors; (c) TDM-DD-QAM with prediction errors.}
	\label{fig_CmpPred} 
	\vspace{-6mm}
\end{figure*}    
  
  We select FRaC, which is also based on mmWave radar, as our baseline for comparison\cite{ref_JSTSP_DingyouMa_IM}.
  This scheme divides a full chirp into equal sub-chirps and assigns them to different activated transmit antennas. Additionally, each sub-chirp undergoes PSK modulation. Therefore, FRaC encodes communication bits from three sources: 1) the selection of activated antennas, 2) the pairing relationship between antennas and sub-chirps, and 3) the PSK modulation of each sub-chirp.
  We adapt the FRaC scheme to the scenario presented in this paper. Accordingly, our DD-QAM scheme also adopts a TDM mode to align with FRaC. Specifically, with our $2\times2$ transmit antenna array, FRaC selects two antennas for each transmission and employs QPSK modulation.
  In this configuration, the antenna selection in FRaC encodes 2~bits, the antenna-sub-chirp pairing encodes 1~bit, and the two QPSK symbols encode 2~bits each, totaling 7~bits per chirp. In contrast, our DD-QAM scheme encodes 9~bits in the delay dimension and 2~bits with QPSK, for a total of 11~bits. The bits modulated in the Doppler dimension are disregarded here as modulation occurs only once per frame. Since the radar sensing front-end for FRaC is similar to ours, their sensing performances are comparable. However, FRaC requires a maximum likelihood estimation detector for bit detection, whereas our method directly demodulates bits from the radar detection results, thus incurring no additional communication complexity overhead.
  
  The SER performance comparison is illustrated in Fig.~\ref{fig_Cmp}, where each curve represents the SER where bits are only modulated in the dimension mentioned in the legend. FRaC achieves favorable communication performance only when modulating bits using QPSK alone. The performance is secondary when encoding information by swapping the antenna-sub-chirp pairings, and worst for antenna selection. If all three dimensions are modulated simultaneously, the bits re nearly impossible to demodulate due to severe errors. This issue arises from the coupling between the phase introduced by the antenna elements and the phase from QPSK modulation, which leads to ambiguity with multiple possible solutions. Furthermore, our transmitter has an antenna spacing of 2$\lambda$, which introduces grating lobes that exacerbate this problem. In contrast, our proposed scheme does not suffer from such ambiguity in demodulation.
  
  In Figs.~\ref{fig_FRaC} and \ref{fig_Mine}, each curve represents the SER where bits are only modulated in the dimension mentioned in the legend. ``All" indicates that all available dimensions are utilized for modulation.
  In the legends for Figs.~\ref{fig_FRaC} and \ref{fig_Mine}, $\Delta$ denotes the modulation interval in the delay dimension, measured in units of delay resolution. ``Pred. Err" refers to the normalized prediction error relative to the IF resolution.
  Prediction error arises from the IF shift induced by the channel. The channel's IF represents the target's delay/range, which is estimated via a tracking process. Since data demodulation relies on this predicted channel state, prediction errors directly impact demodulation performance.  
  As shown in Fig.~\ref{fig_FRaC}, similar to Fig.~\ref{fig_Cmp}, simultaneous modulation in both spatial and phase dimensions results in demodulation failure. When modulating in a single dimension, QPSK achieves a lower SER than spatial modulation. The performance of both schemes deteriorates as the prediction error increases.
  In Fig.~\ref{fig_Mine}, the demodulation performance of both delay modulation and QPSK likewise degrades with increasing prediction error. However, for delay modulation, the symbol errors can be mitigated by increasing the delay modulation interval ($\Delta$). This is an advantage over FRaC, which cannot resolve errors from prediction inaccuracies without resorting to additional channel coding.
  
  \begin{figure}[!t]
  		\vspace{-7mm}
  	\centering
  	\subfigure[]{
  		\begin{minipage}[t]{0.5\linewidth}
  			\centering
  			\includegraphics[width=1.5in]{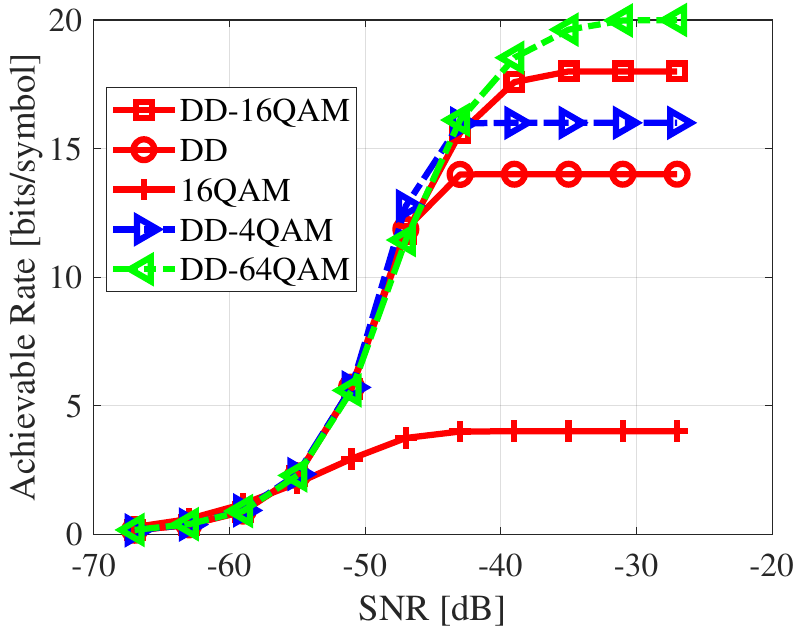}
  			\label{fig_DDM} 
  		\end{minipage}%
  	}%
  	\subfigure[]{
  		\begin{minipage}[t]{0.5\linewidth}
  			\centering
  			\includegraphics[width=1.5in]{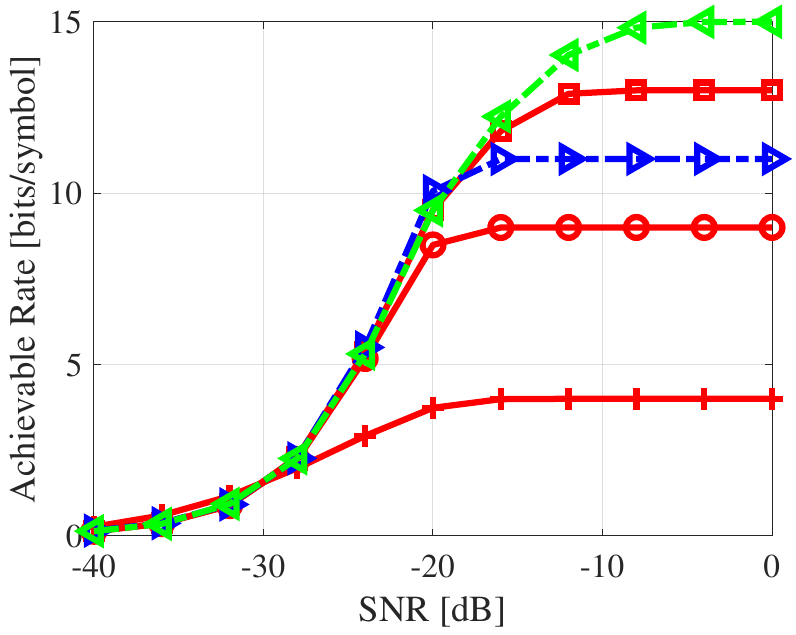}
  			\label{fig_TDM} 
  		\end{minipage}%
  	}%
  	\centering
  	\vspace{-2mm}	
  	\caption{Achievable rate of DD-QAM: (a) DDM;
  		(b) TDM.}
  	\label{fig_rate} 
  	\vspace{-6mm}
  \end{figure}
\vspace{-4mm}
\subsection{Achievable Rate Evaluation}  
Fig.~\ref{fig_DDM} illustrates achievable rates of the proposed DDM scheme in this paper, while Fig.~\ref{fig_TDM} presents those of the TDM scheme, which can be directly extended from the proposed method.
The achievable rate of conventional QAM modulation alone is the lowest.
By introducing the proposed DD-domain modulation, the achievable rate improves by 14~bits/symbol in the DDM scheme and by 9~bits/symbol in the TDM scheme.
The 5~bits/symbol higher achievable rate in the DDM compared to the TDM scheme originates from additional modulation in the Doppler-domain.
Although the achievable rate of the DDM scheme appears higher than that of the TDM scheme in Fig.~\ref{fig_rate}, the symbol rate of TDM will actually be $N_c$ times that of DDM.
This is because, in the TDM scheme, both delay-domain and QAM data can be modulated in each chirp.
Consequently, the achievable data rate of TDM significantly surpasses that of DDM.
However, since the DDM scheme transmits signals from all antennas simultaneously, it achieves $10\log_{10}(N_{\text{tx}})$~dB SNR gain advantage over TDM.
As shown Fig.~\ref{fig_DDM}, with four transmit antennas, the DDM scheme exhibits an overall leftward shift of $27$~dB compared to the TDM scheme.
The additional $21$~dB gain results from the accumulation of $N_c$ chirps within a single frame.
\vspace{-4mm}
\subsection{Tracking Performance Evaluation}\label{S7.3}

  Fig.~\ref{fig_RT} shows the tracking process of the AV. In Fig.~\ref{fig_RT_4D}, the green line represents the true parameter values, the red markers represent the measured values obtained using Algorithm~\ref{alg_ParaEst}, and the blue markers represent the results fused through Algorithm~\ref{alg_EKF_RT}.
 The AV exhibits off-grid estimation capabilities for distance (delay), AoA and radial-velocity (Doppler). Moreover, the AV successfully establishes tracking for new vehicles entering its FoV without disrupting its ongoing tracking process for the PV. 
  This showcases the multi-target detection and tracking capabilities of Algorithm~\ref{alg_EKF_RT}, which is suitable for dynamic scenarios.
  Fig.~\ref{fig_RT_Orientation} shows that the velocities in the $x$- and $y$-directions as well as the orientation can be obtained.
  In contrast to the conventional schemes \cite{ref_TAES_HorizontalVel}, the results of our scheme indicate that additional parameters or features of targets can be obtained by exploiting temporal information in multiple frames.
  Note that when the PV and the vehicle 3 are not in the FoV of the AV, their true parameters with respect to the AV are still plotted to show their relative motion with the AV.

  Similarly, Fig.~\ref{fig_CT} shows the tracking process of the PV. The perspective in Fig.~\ref{fig_RT} is from the AV, while the perspective in Fig.~\ref{fig_CT} is from the PV, both depicting the same underlying physical process.
  The PV can obtain the off-grid estimation of the AV's parameters with respect to itself.

\vspace{-3mm}
\section{Conclusions}\label{Sec_Conclusion}
  This paper has focused on improving communication functionalities upon sensing-centric autonomous vehicles in the mmWave band.
  Our original contribution has been threefold.
  First, in terms of waveform design, we have proposed a DD-QAM-based chirp waveform, enabling efficient data modulation onto delay, Doppler and complex amplitude of transmitted signal.
  Second, for sensing, we have proposed the beacon frame-aided 4D-parameter estimation scheme, achieving quasi-off-grid estimation of delay and Doppler. Furthermore, leveraging temporal information, we have proposed an EKF-based 5D-parameter estimation scheme for the AV. This scheme supports truly dynamic off-grid estimations and additionally offers tangential-velocity and orientation estimations.
  Third, for communication, we have proposed a dual-compensation-based demodulation and tracking scheme for the PV. This scheme leverages temporal information to facilitate data demodulation without compromising sensing functionality.
  Although our design emphasizes sensing, it uncovers a promising path toward communication rate improvement. 
  The achievable bit rate may scale with the number of chirps per frame, suggesting a valuable direction for future research.
\vspace{-2mm}
\bibliographystyle{IEEEtran}
\bibliography{ref}


\vfill

\end{document}